\RequirePackage{lineno}

\documentclass[twocolumn,showpacs,preprintnumbers,superscriptaddress,amsmath,amssymb,floatfix]{revtex4}

\usepackage{psfig}
\usepackage{subfig}
\usepackage{float}
\usepackage{graphicx}
\usepackage{dcolumn}
\usepackage{bm}
\usepackage{epstopdf}
\usepackage{color}
\usepackage{lineno}

\begin{document}


\title{Studies of di-jets in Au+Au collisions using angular correlations with respect to back-to-back leading hadrons } 


\affiliation{AGH University of Science and Technology, Cracow, Poland}
\affiliation{Argonne National Laboratory, Argonne, Illinois 60439, USA}
\affiliation{University of Birmingham, Birmingham, United Kingdom}
\affiliation{Brookhaven National Laboratory, Upton, New York 11973, USA}
\affiliation{University of California, Berkeley, California 94720, USA}
\affiliation{University of California, Davis, California 95616, USA}
\affiliation{University of California, Los Angeles, California 90095, USA}
\affiliation{Universidade Estadual de Campinas, Sao Paulo, Brazil}
\affiliation{Central China Normal University (HZNU), Wuhan 430079, China}
\affiliation{University of Illinois at Chicago, Chicago, Illinois 60607, USA}
\affiliation{Cracow University of Technology, Cracow, Poland}
\affiliation{Creighton University, Omaha, Nebraska 68178, USA}
\affiliation{Czech Technical University in Prague, FNSPE, Prague, 115 19, Czech Republic}
\affiliation{Nuclear Physics Institute AS CR, 250 68 \v{R}e\v{z}/Prague, Czech Republic}
\affiliation{University of Frankfurt, Frankfurt, Germany}
\affiliation{Institute of Physics, Bhubaneswar 751005, India}
\affiliation{Indian Institute of Technology, Mumbai, India}
\affiliation{Indiana University, Bloomington, Indiana 47408, USA}
\affiliation{Alikhanov Institute for Theoretical and Experimental Physics, Moscow, Russia}
\affiliation{University of Jammu, Jammu 180001, India}
\affiliation{Joint Institute for Nuclear Research, Dubna, 141 980, Russia}
\affiliation{Kent State University, Kent, Ohio 44242, USA}
\affiliation{University of Kentucky, Lexington, Kentucky, 40506-0055, USA}
\affiliation{Institute of Modern Physics, Lanzhou, China}
\affiliation{Lawrence Berkeley National Laboratory, Berkeley, California 94720, USA}
\affiliation{Massachusetts Institute of Technology, Cambridge, MA 02139-4307, USA}
\affiliation{Max-Planck-Institut f\"ur Physik, Munich, Germany}
\affiliation{Michigan State University, East Lansing, Michigan 48824, USA}
\affiliation{Moscow Engineering Physics Institute, Moscow Russia}
\affiliation{National Institute of Science and Education and Research, Bhubaneswar 751005, India}
\affiliation{Ohio State University, Columbus, Ohio 43210, USA}
\affiliation{Old Dominion University, Norfolk, VA, 23529, USA}
\affiliation{Institute of Nuclear Physics PAN, Cracow, Poland}
\affiliation{Panjab University, Chandigarh 160014, India}
\affiliation{Pennsylvania State University, University Park, Pennsylvania 16802, USA}
\affiliation{Institute of High Energy Physics, Protvino, Russia}
\affiliation{Purdue University, West Lafayette, Indiana 47907, USA}
\affiliation{Pusan National University, Pusan, Republic of Korea}
\affiliation{University of Rajasthan, Jaipur 302004, India}
\affiliation{Rice University, Houston, Texas 77251, USA}
\affiliation{Universidade de Sao Paulo, Sao Paulo, Brazil}
\affiliation{University of Science \& Technology of China, Hefei 230026, China}
\affiliation{Shandong University, Jinan, Shandong 250100, China}
\affiliation{Shanghai Institute of Applied Physics, Shanghai 201800, China}
\affiliation{SUBATECH, Nantes, France}
\affiliation{Temple University, Philadelphia, Pennsylvania, 19122}
\affiliation{Texas A\&M University, College Station, Texas 77843, USA}
\affiliation{University of Texas, Austin, Texas 78712, USA}
\affiliation{University of Houston, Houston, TX, 77204, USA}
\affiliation{Tsinghua University, Beijing 100084, China}
\affiliation{United States Naval Academy, Annapolis, MD 21402, USA}
\affiliation{Valparaiso University, Valparaiso, Indiana 46383, USA}
\affiliation{Variable Energy Cyclotron Centre, Kolkata 700064, India}
\affiliation{Warsaw University of Technology, Warsaw, Poland}
\affiliation{University of Washington, Seattle, Washington 98195, USA}
\affiliation{Wayne State University, Detroit, Michigan 48201, USA}
\affiliation{Yale University, New Haven, Connecticut 06520, USA}
\affiliation{University of Zagreb, Zagreb, HR-10002, Croatia}

\author{L.~Adamczyk}\affiliation{AGH University of Science and Technology, Cracow, Poland}
\author{G.~Agakishiev}\affiliation{Joint Institute for Nuclear Research, Dubna, 141 980, Russia}
\author{M.~M.~Aggarwal}\affiliation{Panjab University, Chandigarh 160014, India}
\author{Z.~Ahammed}\affiliation{Variable Energy Cyclotron Centre, Kolkata 700064, India}
\author{A.~V.~Alakhverdyants}\affiliation{Joint Institute for Nuclear Research, Dubna, 141 980, Russia}
\author{I.~Alekseev}\affiliation{Alikhanov Institute for Theoretical and Experimental Physics, Moscow, Russia}
\author{J.~Alford}\affiliation{Kent State University, Kent, Ohio 44242, USA}
\author{C.~D.~Anson}\affiliation{Ohio State University, Columbus, Ohio 43210, USA}
\author{D.~Arkhipkin}\affiliation{Brookhaven National Laboratory, Upton, New York 11973, USA}
\author{E.~Aschenauer}\affiliation{Brookhaven National Laboratory, Upton, New York 11973, USA}
% ...

\collaboration{STAR Collaboration}\noaffiliation

\begin{abstract}

Jet-medium interactions are studied via a multi-hadron correlation technique (called ``2+1''), where a pair of back-to-back hadron triggers with large transverse momentum is used as a proxy for a di-jet.
This work extends the previous analysis for nearly-symmetric trigger pairs with the highest momentum threshold of trigger hadron of 5~GeV/$c$ with the new calorimeter-based triggers with energy thresholds of up to 10~GeV and above.
The distributions of associated hadrons are studied in terms of correlation shapes and per-trigger yields on each trigger side. 
In contrast with di-hadron correlation results with single triggers, the associated hadron distributions for back-to-back triggers from central Au+Au data at $\sqrt{s_{NN}}$=200~GeV show no strong modifications compared to d+Au data at the same energy. 
An imbalance in the total transverse momentum between hadrons attributed to the near-side and away-side of jet-like peaks is observed. 
The relative imbalance in the Au+Au measurement with respect to d+Au reference is found to increase with the asymmetry of the trigger pair, consistent with expectation from medium-induced energy loss effects.
In addition, this relative total transverse momentum imbalance is found to decrease for softer associated hadrons.
Such evolution indicates the energy missing at higher associated momenta is converted into softer hadrons.

\end{abstract}

\maketitle




\section{Introduction}

Angular di-hadron correlations with respect to a single charged or neutral high-$p_{T}$ trigger at the center-of-mass energy per nucleon pair $\sqrt{s_{NN}}$= 200~GeV have been found to differ significantly between heavy ion events and more vacuum-like  $pp$ or d+Au collisions.
On the away-side ($\Delta\phi\sim\pi$) of the trigger hadron,  broader correlation distributions and softer transverse momentum ($p_T$) spectra of associated hadrons have been reported for central Au+Au events~\cite{star_corr, phenix_corr}.
In some  associated hadron $p_T$ ranges ($\sim$2 GeV/$c$) the modified away-side no longer resembles the jet-like peak but shows a concave shape near $\pi$.
Novel features have also been discovered in Au+Au data on the near-side (small relative azimuth) of the trigger hadron:  a long-range longitudinal plateau in relative pseudorapidity ($\Delta\eta$), called the ``ridge''~\cite{star_ridge}.
Multiple theoretical models have been proposed to simultaneously explain these structures, including in-medium parton energy loss~\cite{eloss_a,eloss_b,eloss_c,eloss_d,eloss_e,renk_kt} and initial state medium fluctuations leading to higher-order flow components such as the triangular flow ($v_3$)~\cite{flow_v3_theory}.
However, experimental measurements based on 2-particle correlations with respect to high-$p_T$ trigger introduce 
surface bias for the initial hard scattering. Thus the partons and subsequently formed jets on the near- and away-sides can be affected by different underlying physics.
On the other hand,  di-hadron correlations between two  high-$p_T$ particles exhibit jet-like peaks in both near- and away-sides~\cite{star_dijet, phenix_nomodify} with little shape modification from d+Au to central Au+Au, but a strong suppression on the away-side amplitude. 
This observation may be interpreted in two jet-medium interaction scenarios.
One possibility is in-medium parton energy loss followed by in-vacuum fragmentation, which naturally explains the observed high-$p_T$ inclusive hadron suppression, $R_{\rm{AA}}$, defined as a ratio of spectra measured in Au+Au collisions with respect to binary-scaled $pp$ reference~\cite{star_raa,phenix_raa}.
Such scenario can also explain little to no  modifications of near-side peaks in the two-particle correlations ($I_{\rm{AA}}$)~\cite{star_dijet, phenix_nomodify}; however, lack of broadening on the away-side of such correlation functions poses a challenge for this model~\cite{phenix_nomodify}.
Alternatively, there could be  a finite probability for both partons to escape the medium without interactions.  In the model this could be simplified as the ``core/corona'' model, where the dynamics of jets and di-jets in the medium are the same as in vacuum, unless they traverse the ``core'' of the medium where they are fully absorbed. 
In such scenario, the relative high $p_T$ per-trigger yield $I_{AA}$ on the away side, would be expected to be equal to the relative inclusive single high $p_T$ particle yield $R_{AA}$.
Varying the surface bias experimentally could allow differentiating between the models. 

The analysis presented in this paper uses a 3-particle (``2+1'') correlation technique for the data from the Solenoidal Tracker at RHIC (STAR)~\cite{STAR}. 
A pair of back-to-back high-$p_T$ trigger particles is used as proxies for di-jet axis, and angular correlation of  lower-$p_T$ charged hadrons with respect to back-to-back trigger pair is considered. 
This technique was first introduced in the previous STAR  publication~\cite{short_2plus1_paper}. In the early work, the primary trigger (T1) and its back-to-back secondary trigger partner (T2) had similar kinematic thresholds of  ($p_T^{\rm{T1}} > 5$~GeV/$c$ and $p_T^{\rm{T2}} > 4$~GeV/$c$).
Following previous works, the near-side is defined as $\eta-\phi$ region close to trigger T1, and away-side  is the $\eta-\phi$ space close to the away-side trigger T2.
It has been found, that such di-jet-like correlations from central Au+Au collisions are similar in both shape and magnitude to those observed in d+Au events at the same incident energy on both near- and away-sides. This suggested a strong surface (tangential) bias of selected di-jets~\cite{short_2plus1_paper}.

In this paper the analysis is further extended to higher primary trigger thresholds. We attempt to use the asymmetry in the energy of the two back-to-back triggers as a tool for changing the surface bias. The expectation is that such asymmetry can partially arise from a longer path length that the away-side parton must travel in the medium, thus  producing differences in the balance of final jet energies.
To control the degree of the surface bias we vary the relative balance between the energies of the primary trigger T1 (near-side) and its back-to-back trigger partner T2 on away-side. For the most asymmetric trigger pair selection in this work the primary trigger has more than twice the energy of its back-to-back partner. 
The correlation functions and the spectra of associated charged particles around each trigger within $\Delta\phi<0.5$ and $|\Delta\eta|<0.5$ are measured and compared between Au+Au and d+Au reference data.

The total transverse momentum of each side of a di-jet is then calculated by summing the $p_T$ for all associated charged hadrons plus the trigger $E_T$ or $p_T$. The di-jet energy imbalance, $\Delta (\Sigma E_T) $, is calculated as difference between the total transverse momentum  between the same-side and away-side jet-like peaks.
The absolute value of the imbalance could be affected by the kinematic selection of the trigger pair unrelated to the jet-medium interactions, for example,  by the $k_T$ effect. To provide quantitative assessment of jet-medium interaction effects and allow discriminating of the theoretical models, the relative di-jet energy imbalance between Au+Au and d+Au data is more informative. 

\section{Data Sets}

In this work the  d+Au and Au+Au collisions at $\sqrt{s_{NN}}$=200~GeV recorded by the STAR collaboration in Runs 2007 and 2008  are analyzed, extending analysis of the previous, smaller, samples from Runs 2003 and 2004. Earlier measurement has been previously reported in~\cite{short_2plus1_paper}; we will use these results here for comparison. The details of new datasets and analysis selections are discussed below.
For a more uniform detector acceptance only  events with a primary vertex position $V_Z$ within 25 cm of the center of the STAR Time Projection Chamber (TPC) along the longitudinal beam direction were used in the analysis.
Run 2007 provided 74M Au+Au minimum-bias (MB) events. These MB events were selected by requiring at least one hit in the Vertex Position Detectors (VPD) located on each side of the TPC 4.5~m away from the nominal interaction point. A subset of such events with additional requirement of at least one high-energy tower (with transverse energy $E_{T} >5.75$ GeV)  in the Barrel Electromagnetic Calorimeter (BEMC) is  referred as the ``high-tower trigger data''. 
Run 2008 collisions provided 6M d+Au high-tower triggers (with $E_T >4.3$ GeV) and 46M VPD minimum-bias triggered events.
To maximize any medium effects, this work focuses on the analysis of 0-20\% most central Au+Au events. 

The BEMC clusters are each built from a set of closely neighboring calorimeter towers and shower-max detector strips with appropriate shape and quality cuts.
BEMC cluster with the highest transverse energy of at least 8~GeV was selected as the primary trigger (T1). The triggered events were grouped into two different bins with $E_{T} \in[8, 10]$~GeV and $E_{T} \in [10, 15]$~GeV.
A highest-$p_T$ charge particle  in the back-to-back azimuthal region ($|\phi^{\rm{T1}} - \phi^{\rm{T2}} - \pi| < 0.2$) is selected as the back-to-back partner trigger (T2) with kinematic requirement of $4 < p_T^{\rm{ T2}} < 10$ GeV/$c$. 
This width of 0.2 is an approximate  width of the away-side  peak in the high-$p_T$ trigger-trigger correlation~\cite{star_dijet},  ensuring a good balance between the di-jet purity and signal-to-background ratio.  For proper primary trigger designation it is also required that the momentum of the  secondary trigger is less than that of the primary one. 
The new data sets, especially benefiting from implementation of triggering capabilities of the  STAR BEMC detector,  allowed for significant improvement of the kinematic reach of this study, and thus larger asymmetry span in the back-to-back trigger partner selection.  The primary trigger T1 threshold has been moved up by a factor of  two while still maintaining a good energy resolution.
The correlation is constructed with all other charged hadrons in an event. These associated particles are required to have transverse momentum  $1.0 < p_T^{\rm{assoc}} <10$ GeV/$c$, and are divided into  groups with different lower thresholds, $[1,10]$, $[1.5,10]$, $[2,10]$, and $[2.5,10]$~GeV/$c$. 
The kinematic selection for associated hadrons is intended to cover the region where the broadening of the away-side correlation and the near-side ridge were previously reported in the 2-particle correlation measurements~\cite{star_corr, star_ridge}.

The associated charged particles reconstructed by TPC are required to pass a set of track quality cuts.
Mid-rapidity $|\eta|<1.0$ tracks selected are required to have at least 20 fit points for good momentum resolution,  and a distance of closest approach (DCA) to the primary vertex of less than 1 cm.
A fiducial cut of $|\eta|<0.75$ is also required for the center of each EMC cluster.
The high-$p_T$ charged particle contamination in the BEMC trigger sample is removed by the following charged veto cut:  a BEMC-cluster trigger is rejected if a charged particle with $p_T >1.5 $ GeV/$c$ is projected to a point within $ |\Delta\phi|<0.015$ and $|\Delta\eta|<0.015 $ of the center of the cluster.

\section{Analysis Method}

\subsection{Building the Correlation Functions}

\begin{figure*}[!hbt]
\subfigure[]{\includegraphics[width=0.32\textwidth]{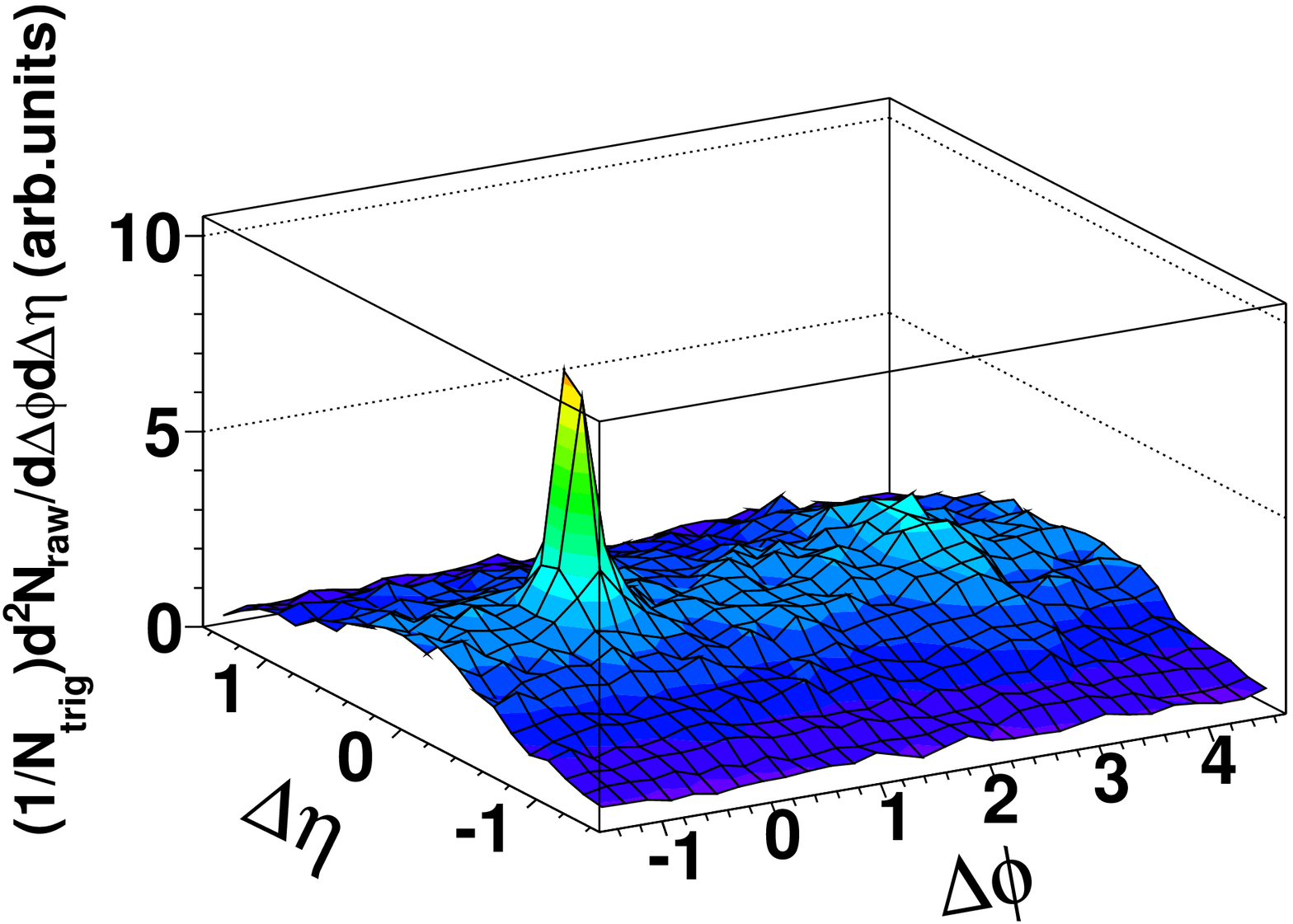}} 
\subfigure[]{\includegraphics[width=0.32\textwidth]{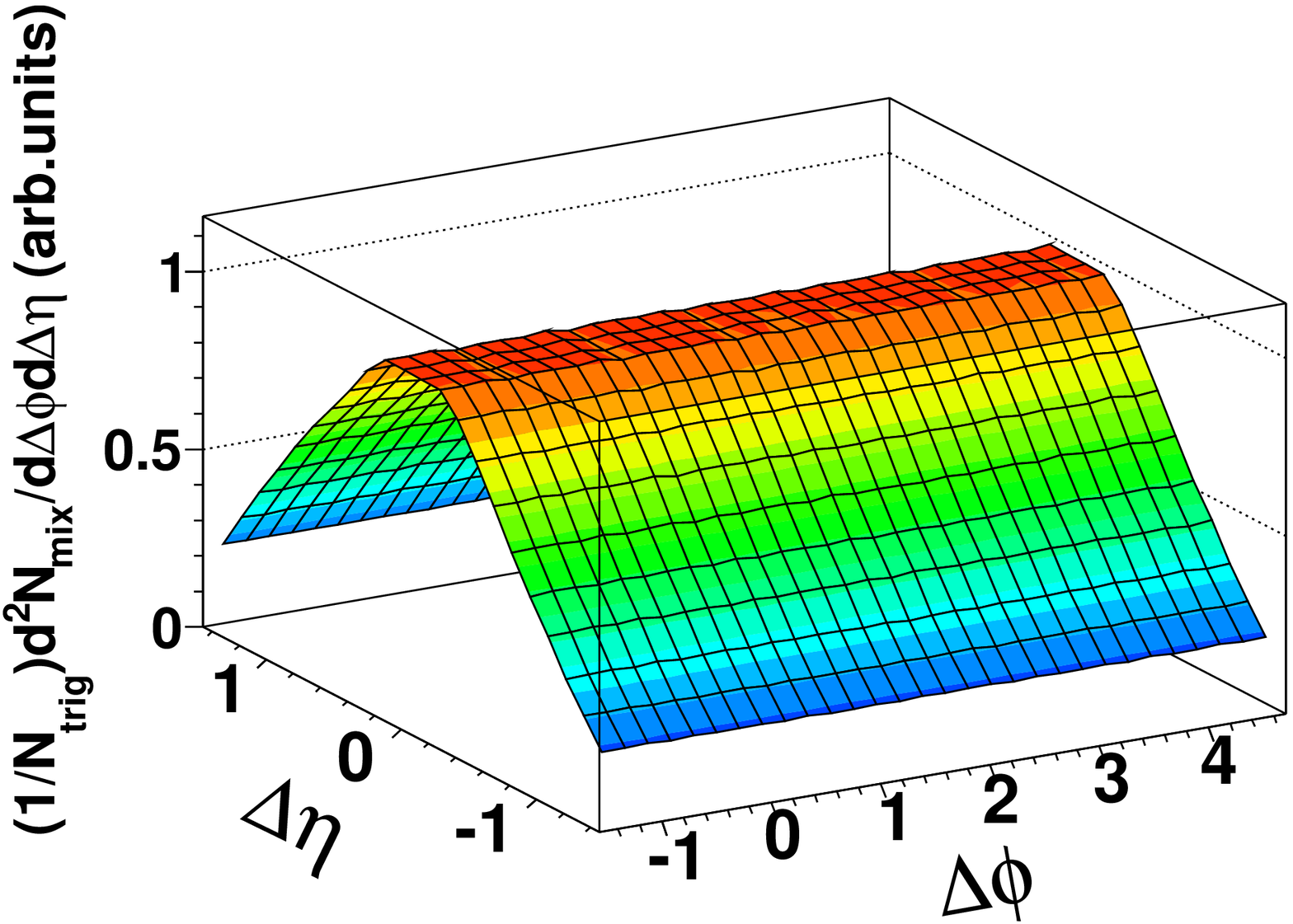} }
\subfigure[]{\includegraphics[width=0.32\textwidth]{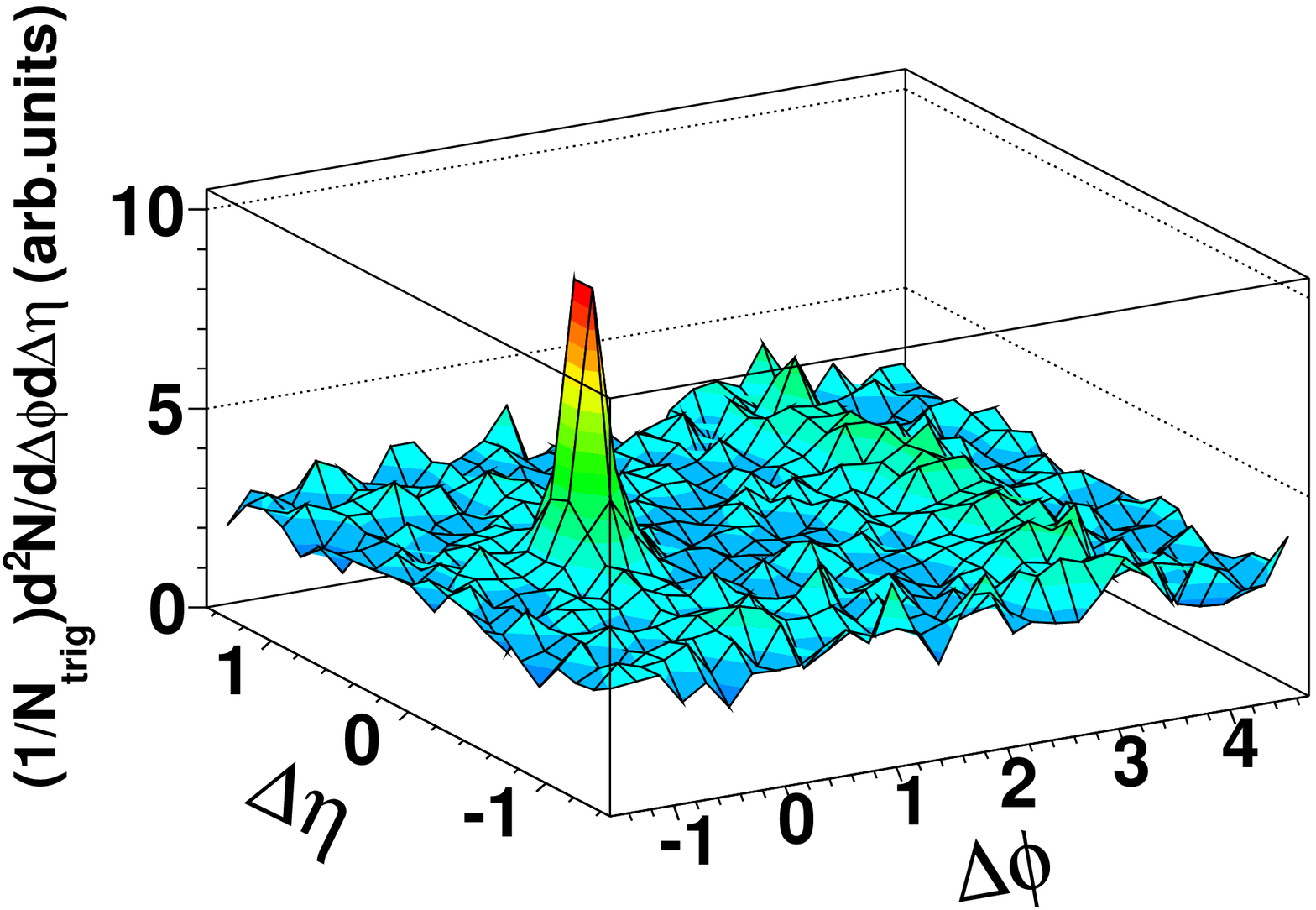} }
\caption{(Color online) (a) A sample of two-dimensional raw correlation function from 200~GeV  0-20\% central Au+Au data before efficiency and acceptance correction. Primary trigger $E_T^{\rm{ T1}} \in$~[8,10]~GeV, secondary trigger $p_T^{\rm{ T2}}\in$~[4,10] ~GeV/$c$, associated hadrons $1.5\leq p_T^{\rm{assoc}}\leq 10$~GeV/$c$. (b) Corresponding two-dimensional mixed-event pair-acceptance correction. (c) Same correlation function corrected for  efficiency and pair acceptance effects. }
\label{figure_sample_mixing_acceptance}
\end{figure*}

The correlation functions studied in this work are defined as

\begin{equation}
\frac{1}{N_{\rm{trig}}}\frac{d^2N}{d\Delta\eta d\Delta\phi}=\frac{1}{N_{\rm{trig}} \rm{\epsilon}_{pair}} (\frac{d^2N_{\rm{raw}}}{d\Delta\eta d\Delta\phi} ) -a_{\rm{zyam}} \frac{d^2N_{\rm{Bg}}}{d\Delta\eta d\Delta\phi}
\label{equation_correlation_function}
\end{equation}
\noindent
where $N_{\rm{trig}}$ is the number of trigger pairs, ${d^2N_{\rm{raw}}}/{d\Delta\eta\,d\Delta\phi}$ is the associated hadron distribution relative to each trigger in correlated (T1, T2) pair, and $\rm{\epsilon}_{pair}$ is a correction factor for single-track efficiency and pair acceptance effects. The ${d^2N_{\rm{Bg}}}/{d\Delta\eta\,d\Delta\phi}$ represents a background term, originating predominantly from randomly associated pairs and correlations due to anisotropic flow. Background scaling factor, $a_{\rm{zyam}}$, is described in the text below.
The tracking reconstruction efficiency is derived by embedding single Monte-Carlo tracks into the real events and reconstructing the combined event.  To account for the $\eta$, $\phi$, $p_T$ and multiplicity dependence of the single track reconstruction efficiency, the correction factor is calculated for each track individually and applied on a track-by-track basis. 
The pair acceptance correction is derived by the mixed-event technique.
For the event mixing all  accepted events with primary vertices $|V_Z|<25$~cm  are divided into ten  5~cm-wide bins. The events are also grouped into 3 centrality sub-classes, corresponding to 0-5\%, 5-10\%, and 10-20\%.
The triggers are then only mixed with associated particles from minimum bias events of the same $V_Z$ and centrality bin to better reflect acceptance of real events and avoid potential trigger biases.
Because the azimuthal acceptance of the STAR detector is uniform, the trigger pairs are used as a whole when mixing with minimum bias events, without the requirement for a correction to account for acceptance variations with respect to the second trigger direction alone.
The two-dimensional ($\Delta\eta$-$\Delta\phi$)  mixed-event correlation function is scaled such that the highest $\Delta\eta \sim 0$ bin is normalized to unity (see e.g. Fig.~\ref{figure_sample_mixing_acceptance}~b). The raw correlation function ($N_{\rm{raw}}$) is then divided by this normalized mixed-event distribution, shown in Fig.~\ref{figure_sample_mixing_acceptance}.

\subsection{Background Subtraction}

The background term $\frac{d^2N_{\rm{Bg}}}{d\Delta\eta d\Delta\phi}$ in Eq.\ref{equation_correlation_function} for each correlation function originates predominantly from random  combinatorics and correlations induced by collective flow.
The shape of this background is described  by multiple components of a Fourier decomposition, with main  contribution in the kinematic region of this analysis from the second-order Fourier component usually associated with elliptic flow ($v_2$).
The multiplicity and $p_T$ dependence of $v_2$ for triggers and associated  particles are obtained from existing STAR measurements~\cite{flow}. To reduce the effect of fluctuations in our $v_2$ estimate we average the results of Event Plane and Four-particle Cumulant methods~\cite{flow-effects}.   
For  $p_T>4$ GeV/$c$ the elliptic flow magnitude is assumed to be constant and at the level  reported in the high-$p_T$ pion flow measurement~\cite{phenix_highpt_pi0_v2}. This assumption is justified since the majority of charged hadrons at these momenta are charged pions~\cite{star_raa,phenix_raa}, and high-$E_T$ clusters used in this work are mainly produced by high-$p_T$ $\pi^{0}$~\cite{phenix_direct_gamma}.
Due to the back-to-back requirement of trigger pair selection, the distribution (apart from efficiency and acceptance effects) is modulated by $f(\Delta\phi) =1+v_2^{2+1}\cos(2\Delta\phi)$, where $v_2^{2+1}$ is resulting flow modulation for three-particle correlation~\cite{fuqiang-model,shinichi-model} given by
\begin{equation}
v_2^{2+1}=\frac{2 v_2^{\rm{T1 \ or \ T2}}v_2^{\rm{assoc}}+2 v_2^{\rm{T2 \ or \ T1}}v_2^{\rm{assoc}}\frac{\sin(2\alpha)}{2\alpha}}{1+2 v_2^{\rm{T1}}v_2^{\rm{T2}} \frac{\sin(2\alpha)}{2\alpha} }. 
\label{equation_v2_modulation}
\end{equation}
\noindent
Here $\alpha$=0.2  is  the half-width of the back-to-back trigger cone. 
%
%
%
The overall background level $a_{\text{zyam}}$ is  estimated with the Zero-Yield at Minimum method~\cite{zyam1,3pPRL,zyam2}.
Each 2-D correlation function within $|\Delta\eta|<1.0$ is first projected on relative azimuth to optimize signal-to-noise ratio and avoid fluctuations at the edges of the TPC acceptance.
Then the zero-yield region for the ``2+1'' correlation  is chosen to be consistently at least 1.3 radians away from both jet-like peaks at $\Delta\phi = 0$~or~$\pi$ .
We note, that this is more than 3$\sigma$ of jet-like peak widths if fit by a Gaussian.
A double Gaussian plus a $v_2$-modulated background fit is also used for the background level estimate to evaluate  systematic uncertainty of the ZYAM method.
The transverse momentum spectra for associated hadrons in the jet peaks is obtained in a similar manner from the $p_T$-weighted correlations selecting the hadrons within 0.5~radians in relative azimuth and 0.5 in relative pseudorapidity of the respective trigger direction.

\begin{figure*}[!ht]
\centering
\subfigure[]{\includegraphics[width=0.32\textwidth]{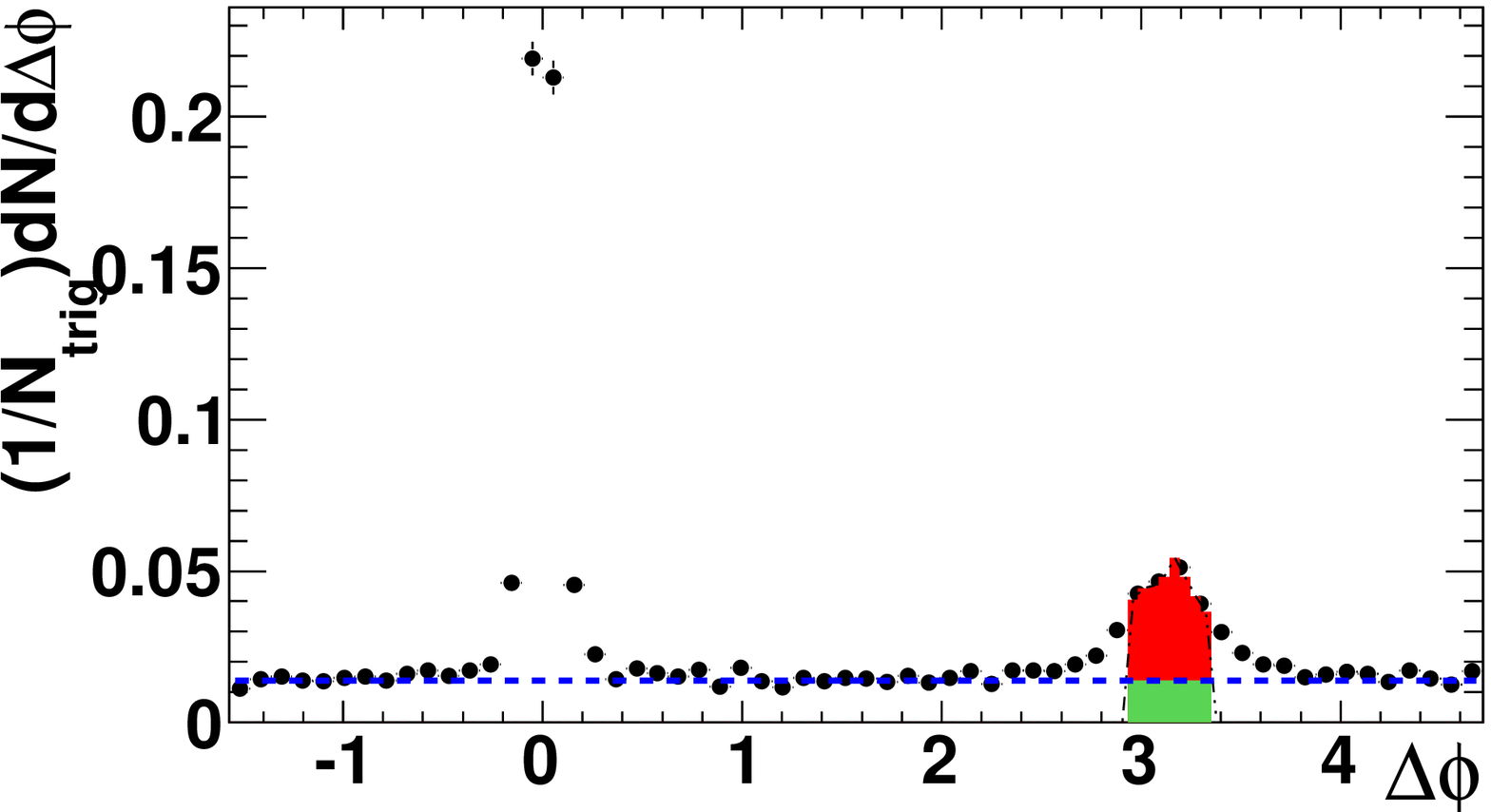}}
\subfigure[]{\includegraphics[width=0.32\textwidth]{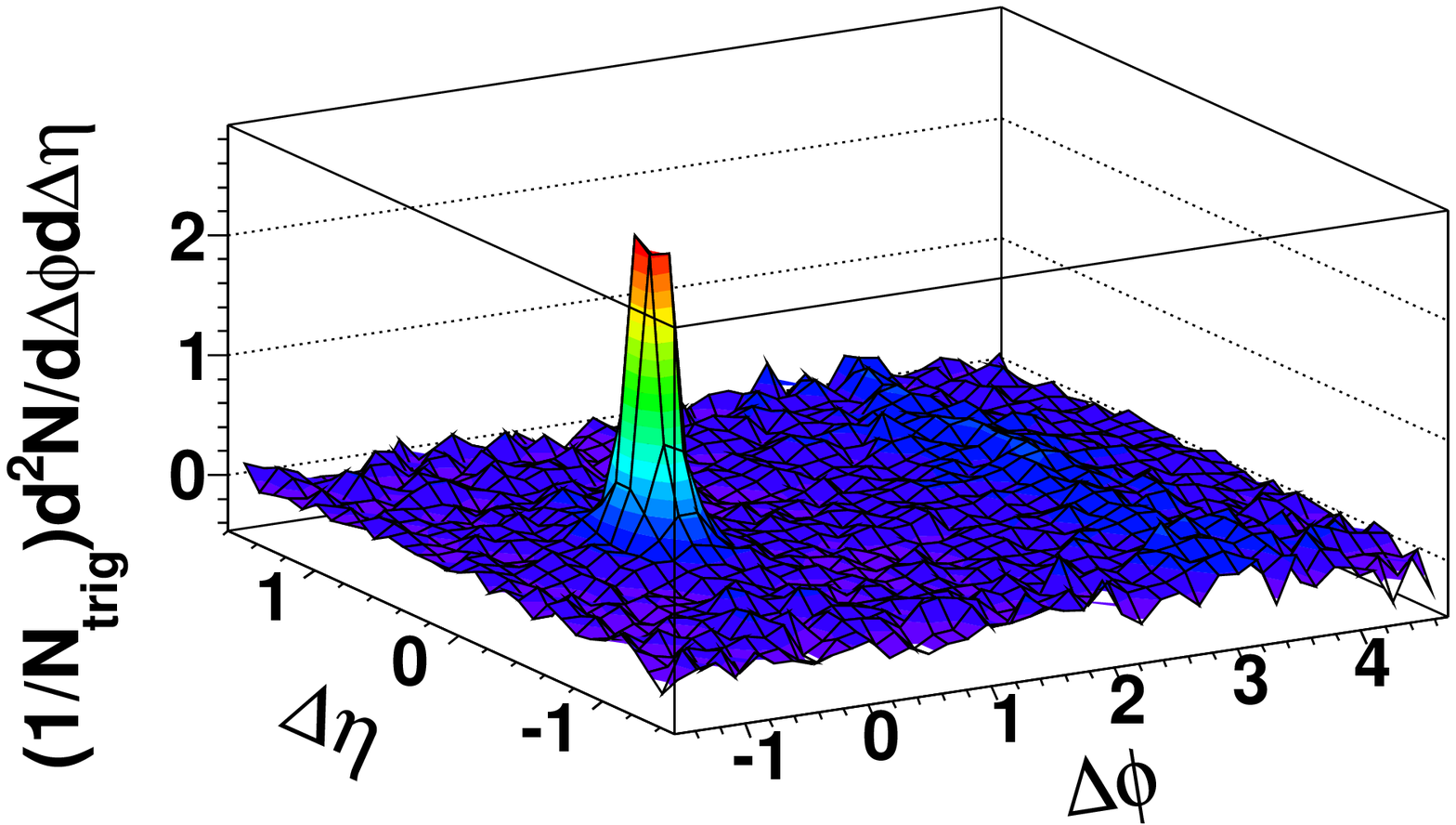}}
\subfigure[]{\includegraphics[width=0.32\textwidth]{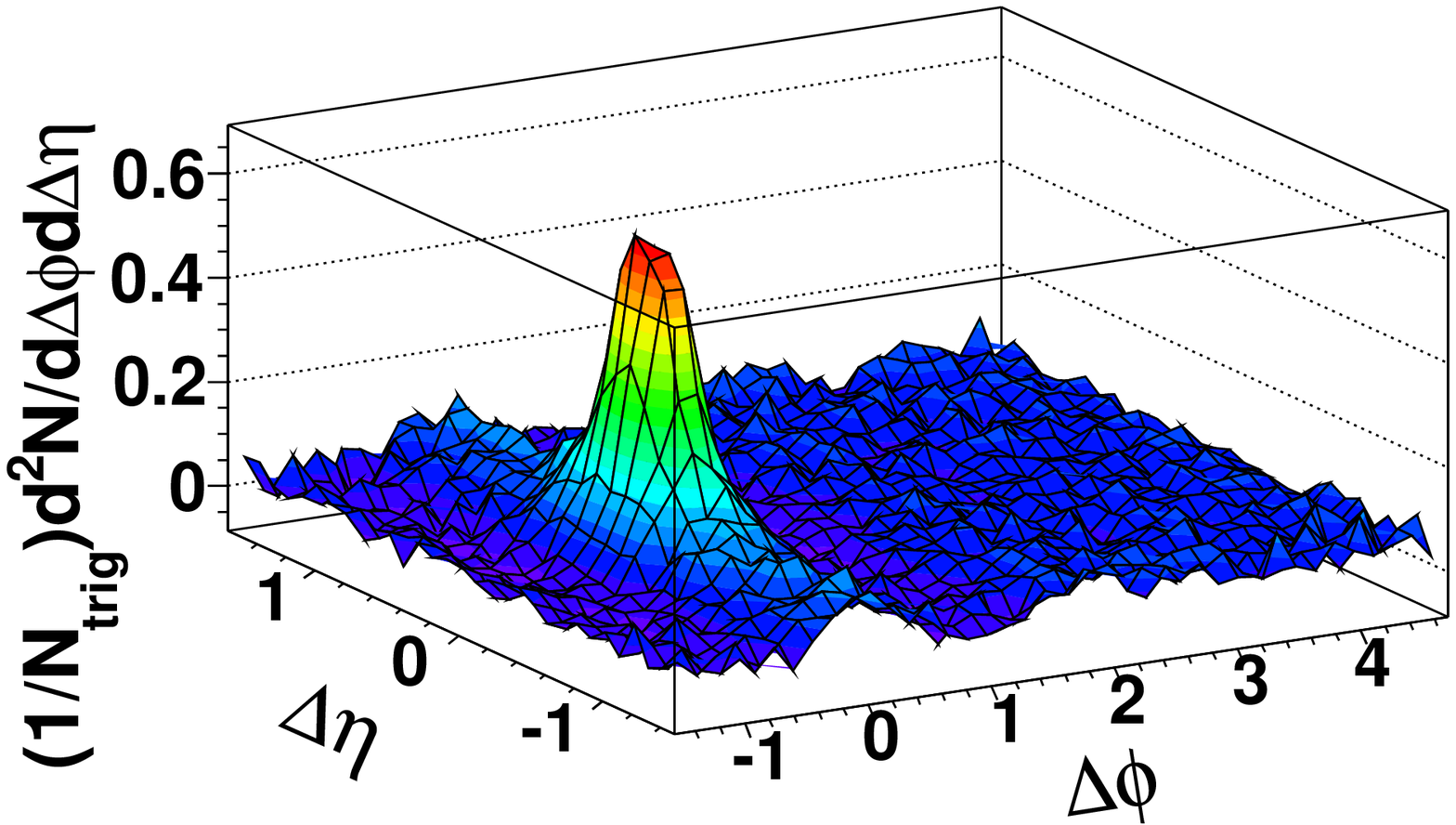}}
\caption{(Color online)(a) Azimuthal trigger-trigger correlation from 0-20\% central 200~GeV Au+Au collisions. Primary trigger $E_T^{\rm{ T1}} \in$~[8,10]~GeV, secondary trigger $p_T^{\rm{ T2}}\in$~[4,10] ~GeV/$c$. 
Dashed line illustrates the ZYAM level.  The contribution from randomly associated pairs for this analysis  is visualized by the green area; from true di-jets - by the red one. (b) Di-hadron correlation for charged hadrons with $1.5\leq p_T^{\rm{assoc}}\leq 10$~GeV/$c$ with respect to triggers matching T1 selection. (c) Di-hadron correlation for charged hadrons with $1.5\leq p_T^{\rm{assoc}}\leq 10$~GeV/$c$ with respect to triggers matching T2 selection. }
\label{figure_sample_2particle_correlation}
\label{figure_trigger_pool}
\end{figure*}

An additional  background term  is related to the randomly associated triggers in the initial selection of the trigger pairs~\cite{3pPRL}.
The signal-to-noise ratio is measured from the  trigger-trigger correlation to estimate the relative  contribution per trigger pair for such random associations.
The correlation contribution due to this background is constructed from two independent 2-particle correlations for T1 and T2 trigger selections separately as in~\cite{short_2plus1_paper}.
The ZYAM method is also applied to these 2-particle correlations, and the zero-yield region is set to be $[0.8, \pi-1.9]$ relative to the trigger, also to avoid both jet-like peaks on near- and away-sides.
An example trigger-trigger correlation to illustrate signal to noise ratio in trigger pairs is presented in Fig.~\ref{figure_trigger_pool}~a.
The dashed line shown in the figure illustrates the  ZYAM level. The number of true di-jet triggers and random pairs is obtained from counting the entries above and below the ZYAM, respectively.
The 2-particle correlation used as an estimate of correlated background contribution corresponding to kinematic selection of first and second triggers are shown in Fig.~\ref{figure_trigger_pool}~b and c, respectively.
We note the known ridge-like  structures  evident in these 2-particle correlations as expected.

\begin{figure*}[!hbt]
\centering
\subfigure[]{\includegraphics[width=0.4\textwidth]{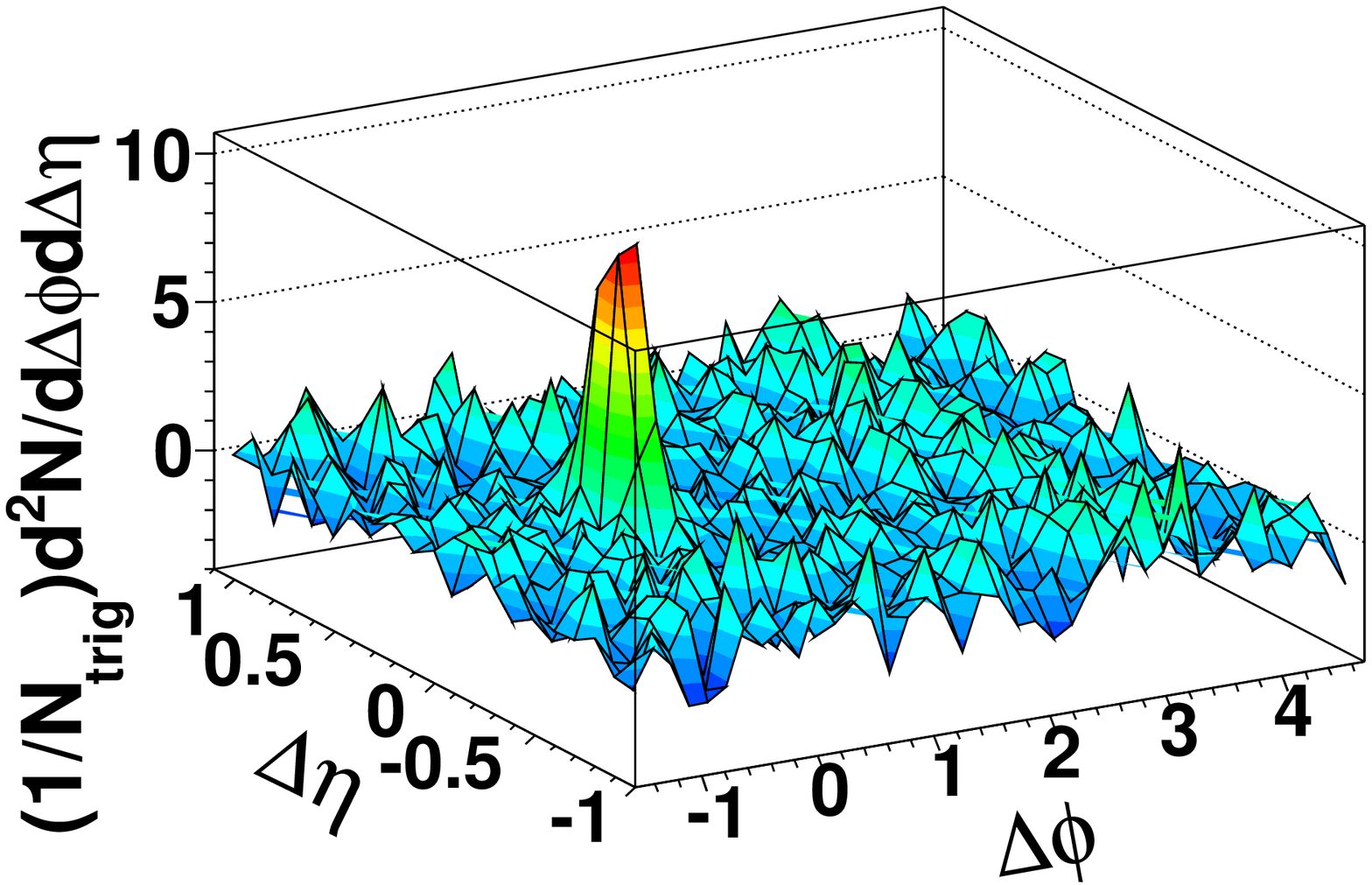}}
\subfigure[]{\includegraphics[width=0.4\textwidth]{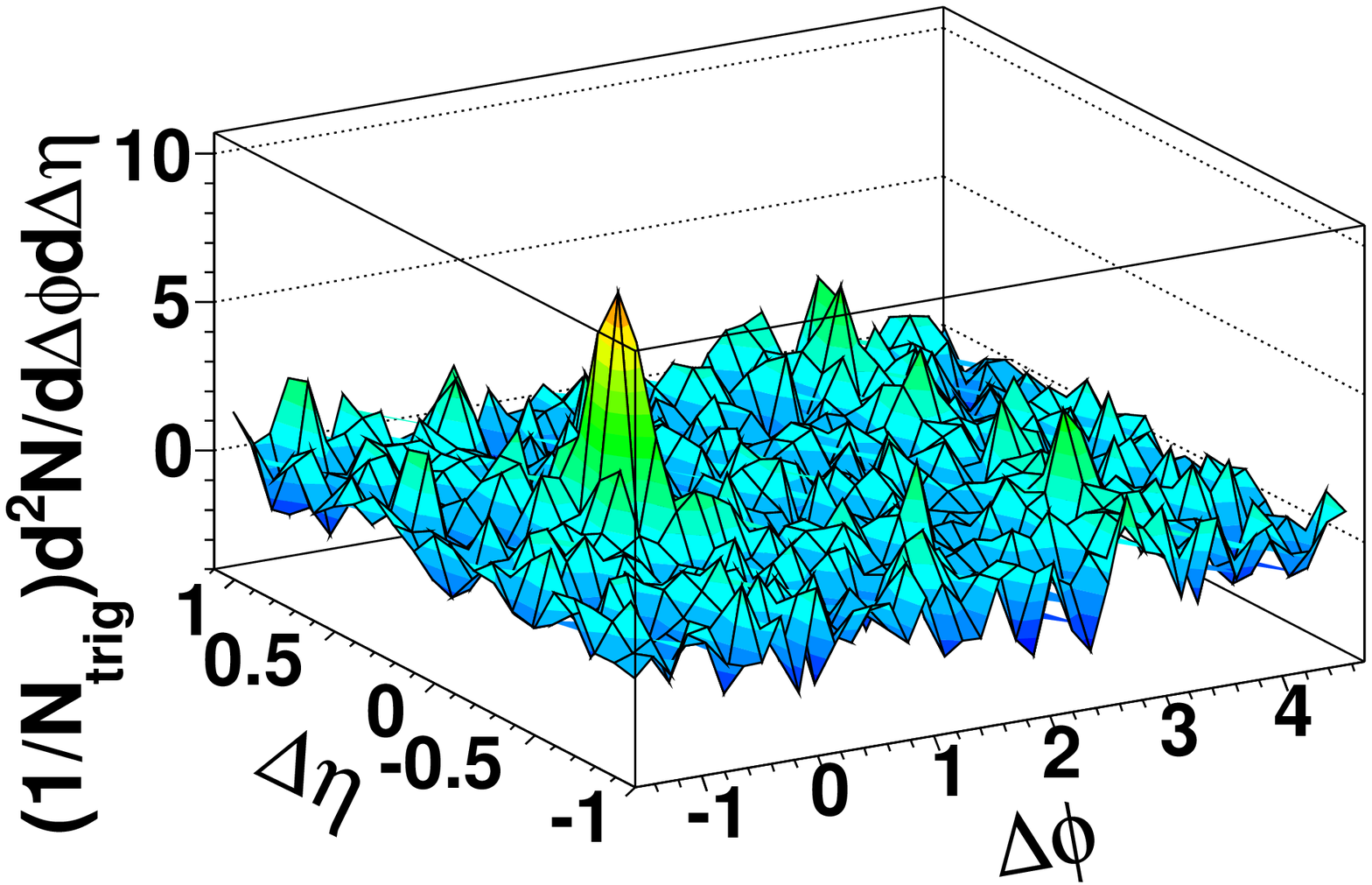} }
\caption{(Color online) Fully corrected  ``2+1'' correlation function from 0-20\% central Au+Au events with respect to the  primary trigger (a) and secondary trigger (b). Primary trigger $E_T^{\rm{ T1}} \in$~[8,10]~GeV, secondary trigger $p_T^{ \rm{T2}}\in$~[4,10]~GeV/$c$, associated hadrons $1.5\leq p_T^{\rm{assoc}}\leq 10$~GeV/$c$. }
\label{figure_sample_2plus1_correlation}
\end{figure*}

A sample set of fully corrected ``2+1'' correlation functions is shown in Fig.~\ref{figure_sample_2plus1_correlation}, with the background induced by uncorrelated trigger pairs and the $v_2$-modulation removed.
In the following sections we will present all the correlation functions measured for the small relative angles only on  both trigger sides to focus on the jet-like peaks. 

An additional test was conducted to estimate possible effects from higher-order flow terms, such as $v_3$, on the correlations.
This was done by including  higher-order flow terms into the fit of  final correlation functions such as in Fig.~\ref{figure_sample_2plus1_correlation}.
We found the effects of $v_3$ to be negligible in the kinetic regime of this analysis.
This result is reasonable as higher-order flow terms are generally explained by the ``soft'' and ``non-jet'' sources; thus the jet-dominant source of our trigger pairs could naturally lead to little or no $v_3$ contribution.
Independent measurements~\cite{phenix_v3_paper} also confirm that the contributions from higher-order  $v_3$ and $v_4$ terms are much smaller than $v_2$ in the trigger $p_T$ region of this paper.
Therefore no corrections for  flow terms other than $v_2$ are included in this analysis.

\section{Results}

\subsection{Correlation Functions and Spectra}
\label{section_correlation_function_spectra_plots}

\begin{figure*}[!hbt]
\includegraphics[width=0.97\textwidth]{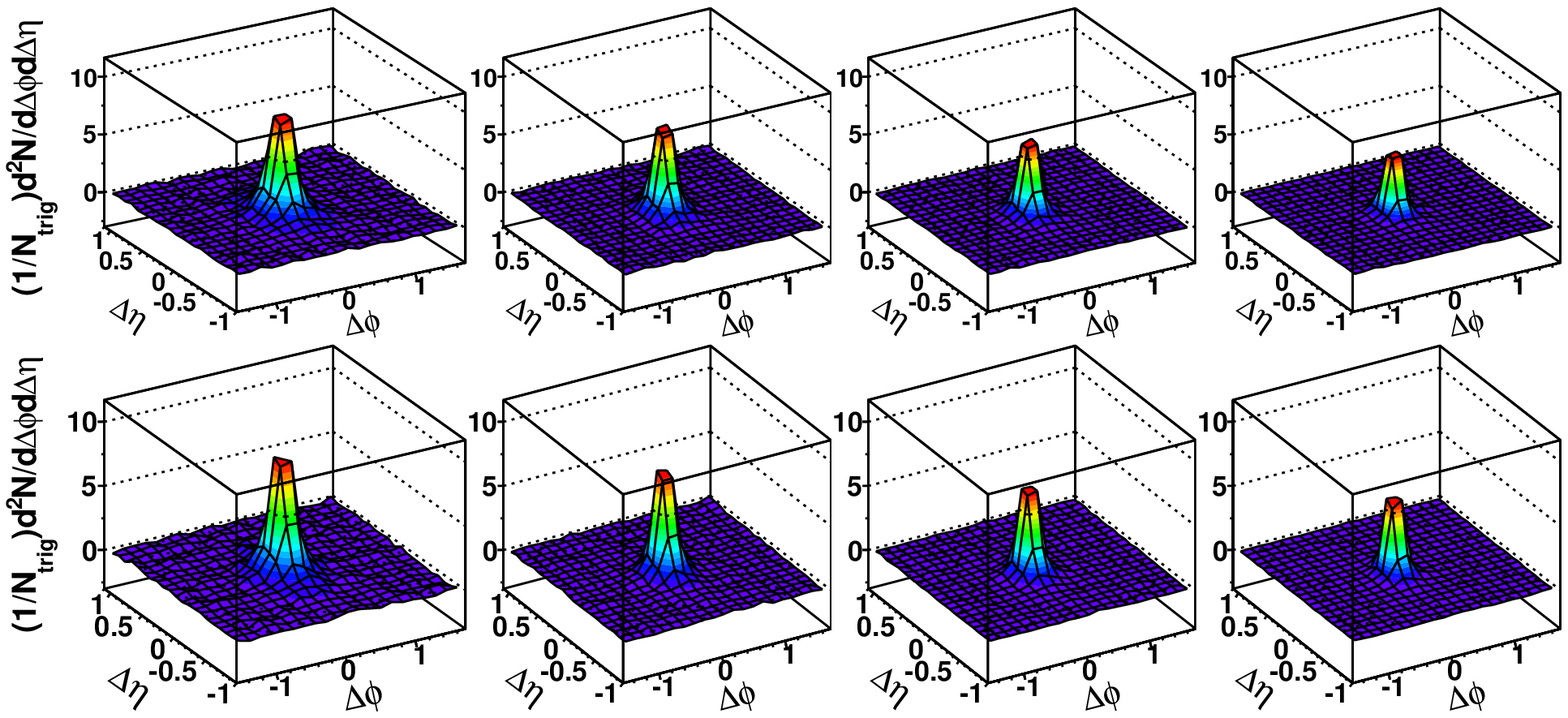}
\caption{(Color online) Near-side associated hadron distributions about primary triggers from 200 GeV d+Au data. The top row shows the correlations for the primary trigger $E_T$ selection of [8,10]~GeV, the bottom row -  for [10,15]~GeV. For all correlations the away-side trigger $p_T$ is in [4,10]~GeV/$c$. From left to right the associated $p_T$ ranges are  [1,10]~GeV/$c$, [1.5,10]~GeV/$c$, [2,10]~GeV/$c$, and [2.5,10]~GeV/$c$.}
\label{figure_run8_2d_near}
\end{figure*}

\begin{figure*}[!hbt]
\includegraphics[width=0.97\textwidth]{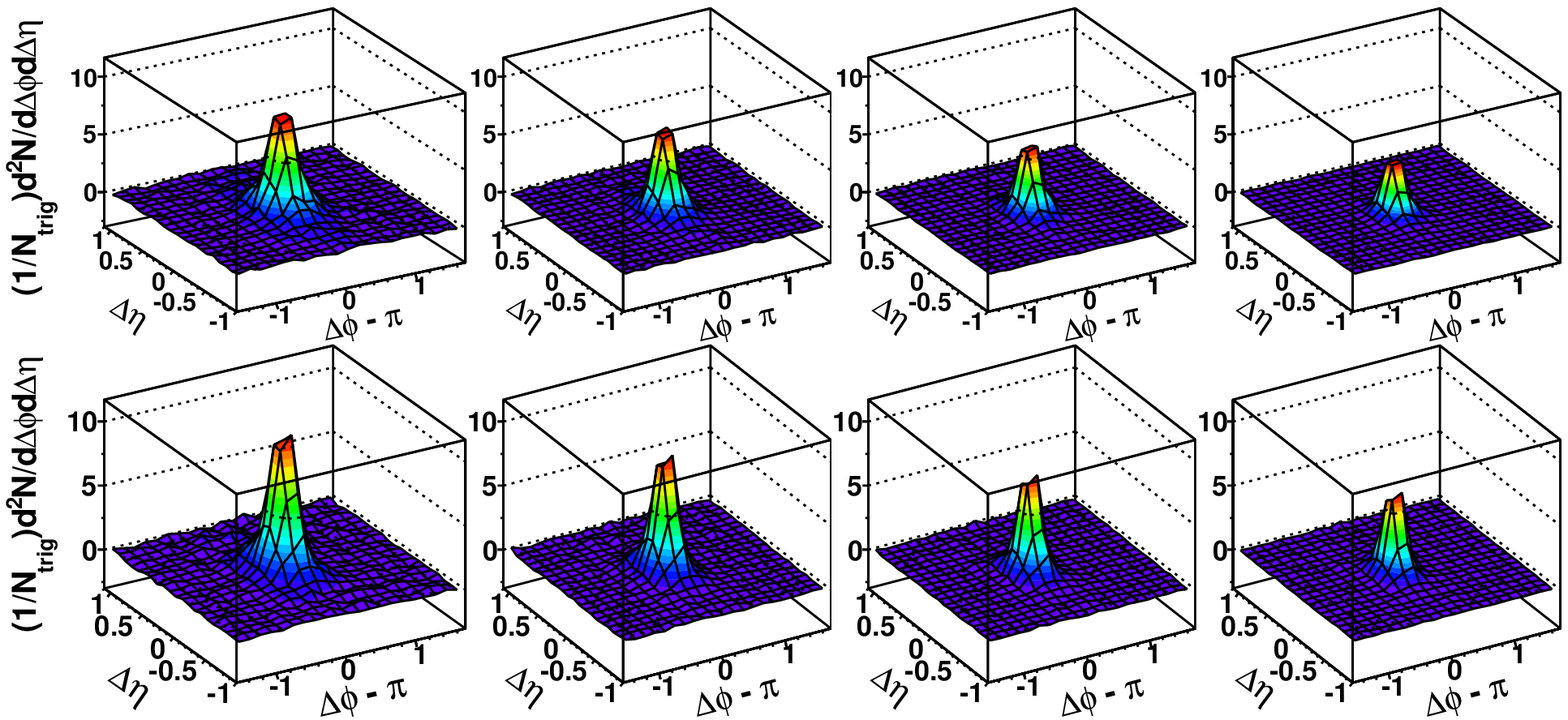}
\caption{(Color online) Away-side associated hadron distributions about secondary triggers from 200 GeV d+Au data. The top row shows the correlations for the near-side trigger $E_T$ selection of [8,10]~GeV, the bottom row -  for [10,15]~GeV. For all correlations the away-side trigger $p_T$ is in [4,10]~GeV/$c$. From left to right the associated $p_T$ ranges are [1,10]~GeV/$c$, [1.5,10]~GeV/$c$, [2,10]~GeV/$c$, and [2.5,10]~GeV/$c$.}
\label{figure_run8_2d_away}
\end{figure*}

\begin{figure*}[!hbt]
\includegraphics[width=0.97\textwidth]{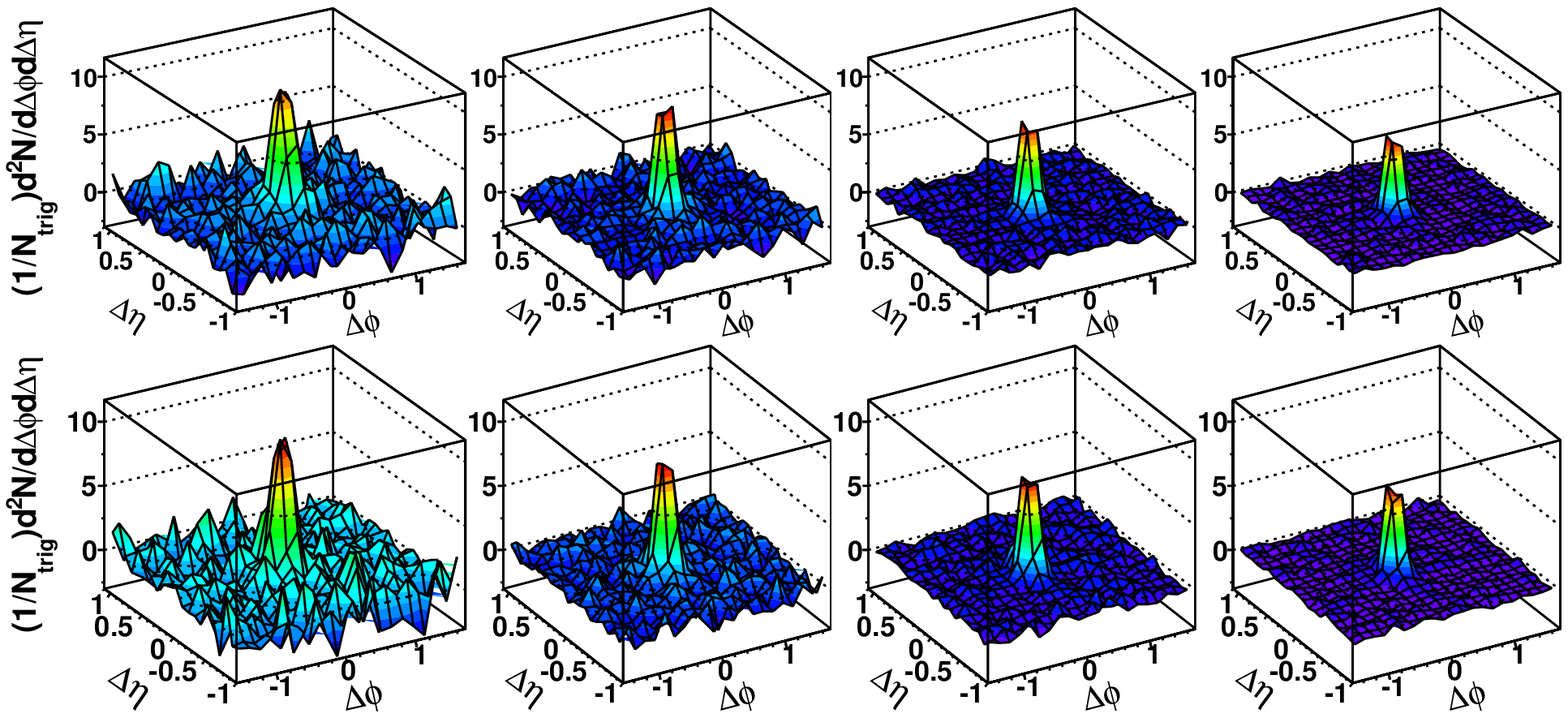}
\caption{(Color online) Near-side associated hadron distributions about primary triggers from 200~GeV 0-20\% central Au+Au data. The top row shows the correlations for the near-side trigger $E_T$ selection of [8,10]~GeV, the bottom row -  for [10,15]~GeV. For all correlations the away-side trigger $p_T$ is in [4,10]~GeV/$c$.
From left to right the associated $p_T$ ranges are [1,10]~GeV/$c$, [1.5,10]~GeV/$c$, [2,10]~GeV/$c$, and [2.5,10]~GeV/$c$.}
\label{figure_run7_2d_near}
\end{figure*}

\begin{figure*}[!hbt]
\includegraphics[width=0.97\textwidth]{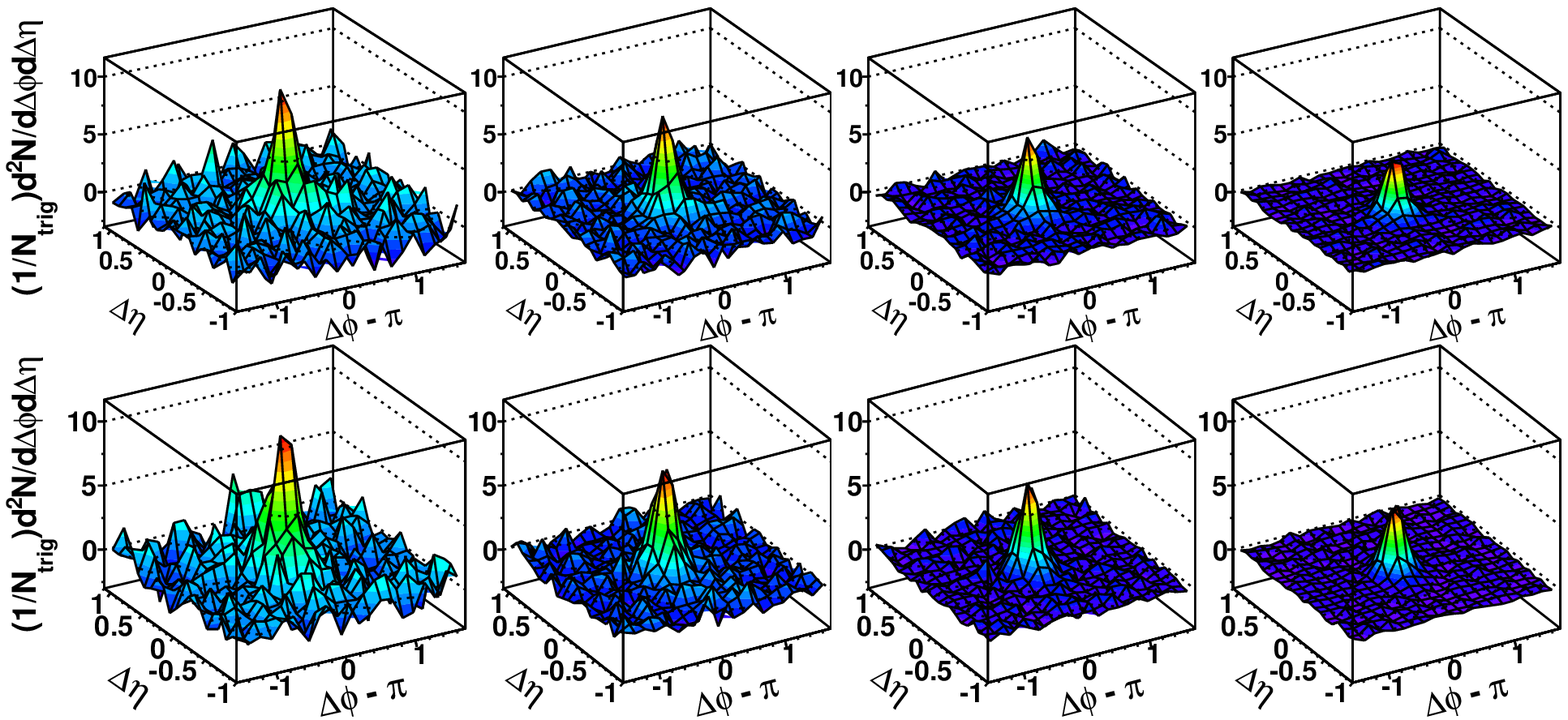}
\caption{(Color online) Away-side associated hadron distributions about secondary triggers from 200~GeV 0-20\% central Au+Au data. The top row shows the correlations for the away-side trigger $E_T$ selection of [8,10]~GeV, the bottom row -  for [10,15]~GeV. For all correlations the away-side trigger $p_T$ is in [4,10]~GeV/$c$. From left to right  the associated $p_T$ ranges are [1,10]~GeV/$c$, [1.5,10]~GeV/$c$, [2,10]~GeV/$c$, and [2.5,10]~GeV/$c$.}
\label{figure_run7_2d_away}
\end{figure*}

In this section we present results obtained from fully corrected correlations for each $p_T$ bin studied for both d+Au and Au+Au datasets. 
The 2-D correlations themselves are shown in Figs.~\ref{figure_run8_2d_near}-\ref{figure_run7_2d_away} for 
primary triggers with transverse energy $8<E_T^{\text{T1}}<10$~GeV and $10<E_T^{\text{T1}}<15$~GeV. The secondary back-to-back trigger for all correlations shown is chosen to have  $4<p_T^{\text{T2}}<10$~GeV/$c$.
Only statistical errors are included in each of these plots.


\begin{figure*}[!hbt]
\includegraphics[width=0.97\textwidth]{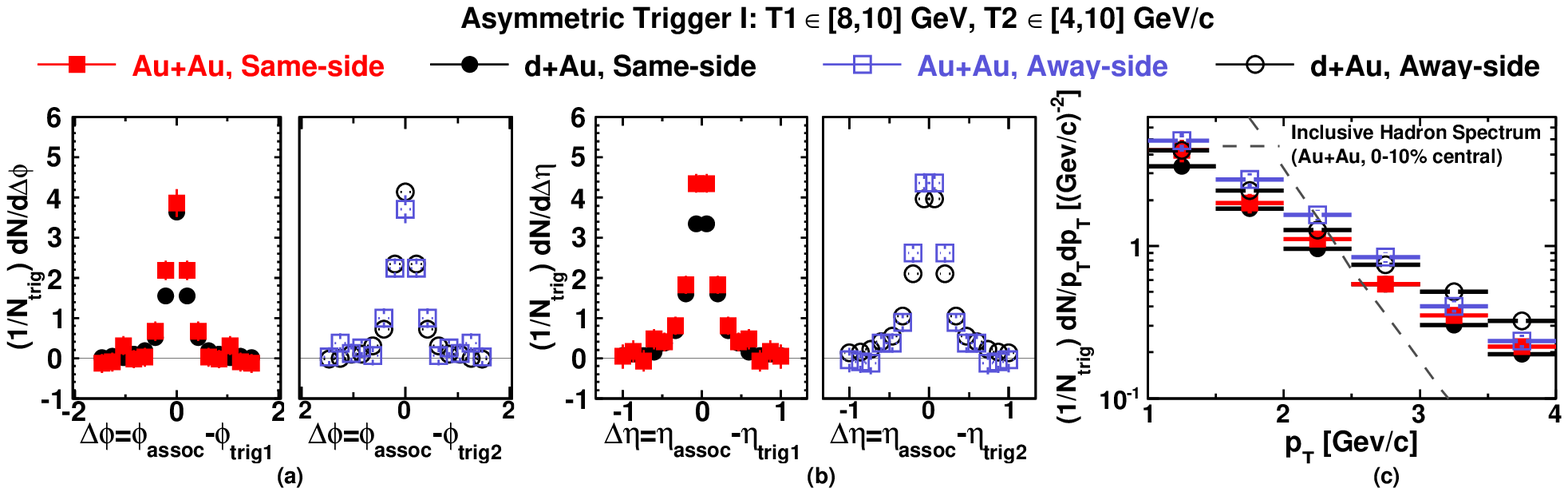}
\caption{(Color online) Projections of 2-D correlation functions on  $\Delta\phi$ ($a$) (with $|\Delta\eta|<1.0$) and $\Delta\eta$ ($b$) (with $|\Delta\phi|<0.7$) for the hadrons associated with their respective triggers (T1 for near-side, T2 for away--side) are shown for d+Au (circles) and central 0-20\% Au+Au (squares).  Errors shown are statistical. The kinematic selection  is as follows: $8<E_T^{\text{T1}}<10$~GeV, $4<p_T^{\text{T2}}<10$~GeV/$c$, $1.0<p_T^{\text{assoc}}<10$~GeV/$c$.
$c)$ Transverse momentum spectral distributions per trigger pair for the near- and away-side hadrons associated with di-jet triggers ($|\Delta\phi|<0.5$, $|\Delta\eta|<0.5$). The inclusive charged hadron distribution from the central 0-10\% Au+Au data is shown for comparison.}
\label{figure_comb_run7_run8_cf_spectra_0}
\end{figure*}

\begin{figure*}[!hbt]
\includegraphics[width=0.97\textwidth]{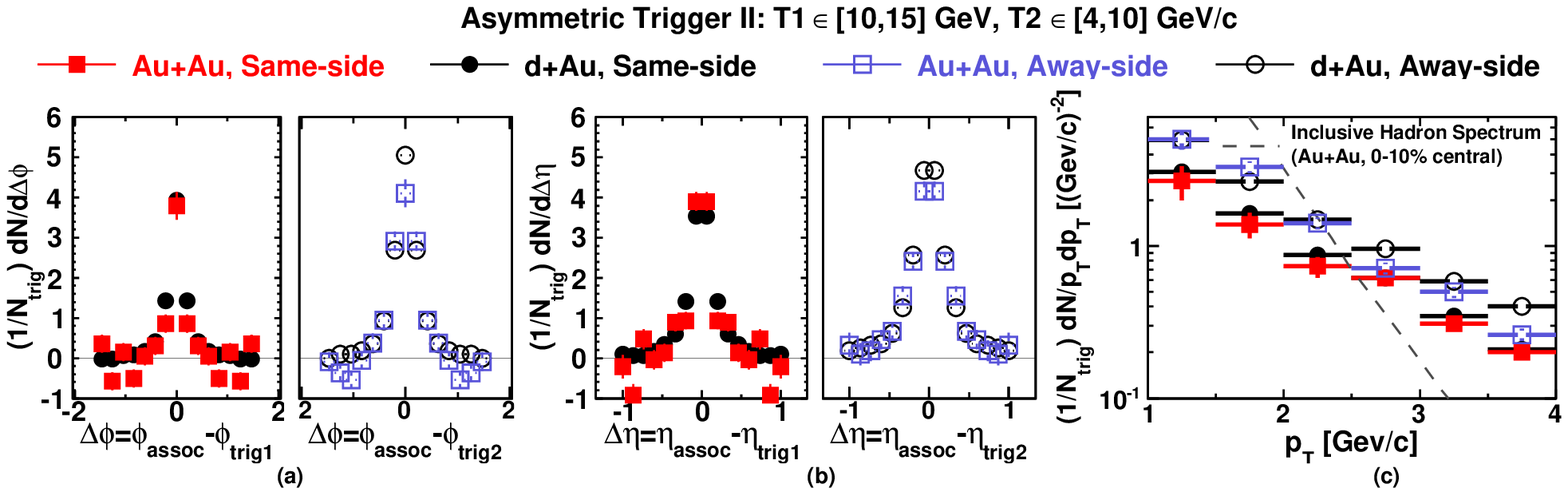}
\caption{(Color online) Projections of 2-D correlation functions on  $\Delta\phi$ ($a$) (with $|\Delta\eta|<1.0$) and $\Delta\eta$ ($b$) (with $|\Delta\phi|<0.7$) for the hadrons associated with their respective triggers (T1 for near-side, T2 for away--side) are shown for d+Au (circles) and central 0-20\% Au+Au (squares) data.  Errors shown are statistical. The kinematic selection  is as follows: $10<E_T^{\text{T1}}<15$ GeV, $4<p_T^{\text{T2}}<10$ GeV/$c$, $1.0<p_T^{\text{assoc}}<10$ GeV/$c$.
$c)$ Transverse momentum spectral distributions per trigger pair for the near- and away-side hadrons associated with di-jet triggers ($|\Delta\phi|<0.5$, $|\Delta\eta|<0.5$). The inclusive charged hadron distribution from the central 0-10\% Au+Au data sample is shown for comparison.}
\label{figure_comb_run7_run8_cf_spectra_4}
\end{figure*}

In Figs. \ref{figure_comb_run7_run8_cf_spectra_0} and \ref{figure_comb_run7_run8_cf_spectra_4} the $\Delta\phi$ and $\Delta\eta$ projections of 2-D correlation plots for the asymmetric back-to-back triggers studied are shown. The associated hadron transverse momentum selection for  projections shown is [1,10]~GeV/$c$, the lowest associated $p_T$ bin studied. Also shown are the transverse momentum spectra  for the associated charged particles within a 0.5x0.5 ($\Delta\eta$ x $\Delta\phi$) area of each trigger, the region containing the dominant part of the jet-like peak.
Only statistical errors are included in each of the plots; the systematic errors  are strongly correlated between the near-/away-side signals of each collision system and are discussed in detail in the text.
We find that for each T1 bin studied, the correlation functions and associated particle spectra from Au+Au events are similar to those observed in the d+Au data in spite of increased asymmetry between T1 and T2 triggers.
The near-side yields of both the  d+Au and the Au+Au data are smaller than their respective away-side yields, and the near-side Au+Au yield slightly drops when $E_T$ of T1 increases from [8,10] to [10,15] GeV. Both observations are possibly due to the increased direct $\gamma$ contamination. This effect will be further discussed later.

\subsection{Systematic Errors}

Several sources of systematic uncertainties have been evaluated as outlined below.
The single track reconstruction efficiency for each centrality and $p_T$ bin was derived as a 2-D function of $\eta$ and $\phi$. The estimated uncertainty for this correction is 5\%  in  each centrality-$p_T$-$\eta$-$\phi$ bin.
This uncertainty is reduced to 1\% when the correlated yields of the near-side are compared to those of the away-side from  the same data sample since the multiplicity-dependent effects cancel.  
The finite statistical uncertainty on the mixed-event correlation leads to a systematic uncertainty in the pair-acceptance correction. This error is estimated to be less than 5\% for each centrality bin.
The uncertainty due to anisotropic flow contribution is estimated from the difference in the $v_2$ results from the Event Plane and Four-particle Cumulant methods. While the fluctuations of $v_2$ itself can be large, the final effect is relatively small due to the evident jet-like peak shapes on both trigger sides. This uncertainty is found to be 5\% in the  central Au+Au. 
This uncertainty is largely correlated between the near- and away-sides. It is not applicable to d+Au events where no $v_2$ modulation is included. 
Since higher-order Fourier terms, such as $v_3$ and $v_4$, have little effect on this analysis, no systematic uncertainties are assigned from these sources.

The systematic uncertainty due to the ZYAM normalization of the background level was estimated by varying  the $\Delta\phi$ range from which the minimum used in the ZYAM method was derived. Other background assumptions, such as a double-Gaussian plus $v_2$-modulated background are also used to estimate the scale of this uncertainty.
The corresponding systematic uncertainty is found to be less than 5\% in this analysis. This uncertainty is strongly correlated between near- and away-sides and cancels in relative side-to-side comparisons. 

The systematic uncertainty due to the  correlated background subtraction from the di-jet sample is determined to be less than 3\% for both d+Au and Au+Au events, and is particularly small in the  high-$E_T$ trigger ranges where signal-to-background ratio is high. This error is also correlated between the near-/away-sides, and is estimated by varying the background normalization for the trigger-trigger correlation in a  manner similar to that used for the ``2+1'' correlation.
Since 2-particle correlations were used for the correction of random combinatorics of trigger pairs, the uncertainty in such di-hadron distributions  arises mainly from the aforementioned sources:  uncertainty in $v_2$ and ZYAM normalization. These uncertainties were evaluated in a similar manner as above and the results are estimated to vary from less than 1\% in d+Au events up to about 5\% in central Au+Au data.

Despite the charged-veto cut, there are still possible contributions to the BEMC cluster trigger energy from low-$p_T$ charged particles.
This contribution is estimated via averaging the energy of those BEMC towers which are far  from each trigger and pass the hot map cuts.
The charged track signal contamination to the high energy tower cluster was estimated to be 2.5\% in central Au+Au and less than 1\% in d+Au.
The overall systematic uncertainty in each  trigger-associated combination bin is less than 15\% after all the contributions mentioned above are summed in quadrature. This is the level of uncertainties important for absolute measurements of the near-/away-side correlation functions or spectra in each collision system. The systematic errors are dominated by sources strongly correlated between near-/away-sides, and will therefore mostly cancel when near- and away-sides are compared. Thus, relative measures, such as energy imbalance, have higher sensitivity to physics effects.

\section{Discussion}

Our previous results~\cite{short_2plus1_paper} have shown similarity of the jet-like peaks not only on the near-side, but also on the away-side of the primary high-$p_T$ trigger between central Au+Au and d+Au data for the  hadrons in the kinematic range of $p_T >$ 1.5 GeV/$c$ associated with the back-to-back high-$p_T$ trigger pair. 
In addition, no evidence of the ``dip'' or ``ridge'' measured in di-hadron correlations with respect to a single trigger of similar kinematic selection is present in the ``2+1'' correlations reported in~\cite{short_2plus1_paper}.
This similarity in the correlation shapes was further supported by the similarities in the associated hadron $p_{T}$ distributions.
Significant softening of the away-side spectra observed in di-hadron correlations~\cite{star_corr}, one of the known indications of energy deposition into the medium, was not evident in these  ``2+1'' data.
A simple  accommodation for the observed differences between di-hadron and ``2+1'' correlation results could be provided by the tangential di-jet emission scenario.  In the presence of very strong energy loss in the core of the medium, only back-to-back parton pairs produced close to the surface and traversing the minimal amount of the medium could be recovered in the analysis. The Au+Au correlation functions would then naturally become similar to those of d+Au.
The ``2+1'' results from symmetric triggers do not, however, exclude path-length dependent energy scenario, where finite in-medium energy loss is followed by in-vacuum fragmentation, for all di-jets.
By increasing the asymmetry in transverse momentum between the two back-to-back trigger hadrons one may control the degree of the surface bias, and thus the in-medium path-length traversed by the parton. The path-length effects on energy loss then could be studied through the imbalance induced by the jet-medium interaction on the final energies of each side of di-jet. 
In this work the asymmetry between the back-to-back triggers is exceeding a factor of two, however, the correlations still show similar jet-like peak shapes and magnitudes from d+Au to central Au+Au collisions for both the near- and away-sides. No evident ridge or dip structure is observed at either near-/away-side in the Au+Au data. 
%
%
We point out that  statistical limitations prevent the complete exclusion of a ridge~\cite{star_ridge}, which could have a very small magnitude in the high-$p_T$ trigger regime studied here.
The absence  of  strong modifications in the  Au+Au correlations relative to d+Au signifies the intensity of jet-medium interactions, as the recovered di-jets still appear to have significant surface bias even for the large trigger asymmetry. We note an increase in the correlation magnitude on away-side with respect to near-side in both Au+Au and d+Au correlations for our most asymmetric trigger pair selection. This increase by itself, as observed in both systems, is likely not related (or not in full)  to the parton/jet energy loss.
Since the trigger pairs pre-selected some di-jet asymmetry,  the total yields and $\Sigma p_T$ of the associated particles from the higher-energy-trigger side must be lower due to energy conservation without any energy loss.
Hence the relative change between Au+Au and the  reference d+Au data is a more relevant measure.
Additionally, since the majority of the  systematic errors are strongly correlated in this analysis  between the near- and away-sides, the measurements of relative differences also have an advantage from experimental point. 
The spectra of associated charged particles for each trigger side from central Au+Au events are  compared to the corresponding distributions from d+Au data  in Fig.~\ref{figure_comb_run7_run8_cf_spectra_0} and Fig.~\ref{figure_comb_run7_run8_cf_spectra_4}.
The  ratios of Au+Au to d+Au spectra  are close to unity and have no prominent $p_T$ dependence, indicating that the softening or suppression, if present, is not nearly as strong as observed in the single particle $R_{\rm{AA}}$.
An additional factor to consider in the comparisons is a possible direct-$\gamma$ contamination of the primary triggers  measured by BEMC clusters, since the near-side yields of direct-$\gamma$ triggers are expected to be close to zero~\cite{phenix_gamma_corr, star_gamma_corr}, potentially reducing the per-trigger measured yields on the near-side.
While the primary triggers from BEMC clusters of $8<E_T^{\rm{T1}}<15$ GeV are dominated by photon pairs from $\pi^{0}$ decays,  grouped into the same BEMC cluster due to  large tower size,
the direct $\gamma$ contamination may not be negligible over the $E_T$ range of this study~\cite{phenix_direct_gamma}.
We expect, however, that  such direct-$\gamma$ contamination would reduce the energy imbalance measured if affecting it at all:  since no in-medium energy loss is expected for direct-$\gamma$, the away-side jet corresponding to such trigger would be relatively higher than that of  $\pi^0$ trigger due to the away-side (reverse) surface bias. 
Meanwhile, the consistent magnitudes of the near-side Au+Au and d+Au correlation functions indicate that the contribution of direct-$\gamma$ triggers is relatively small.
In short, the ``2+1'' correlations can be used as tool for more differential measurements of path-length effects on in-medium energy loss for hard scattered partons.

The difference between the near- and away-side jet energy is used to  quantify the medium effects. The jet energy is  estimated by summing the $E_{T}$ ($p_{T}$) of trigger and  associated charged particles within the 0.5x0.5 ($\Delta\eta$ x $\Delta\phi$) area of each trigger in the kinematic range used in this analysis. The imbalance between the near- and away-sides is measured by the difference of the corresponding energy sums:
$$
\Delta (\Sigma E_T) = (E_T^{\rm{T1}}+ \Sigma p_T^{\rm{assoc, near}}) - (p_T^{\rm{T2}}+ \Sigma p_T^{\rm{assoc, away}})
$$
This energy imbalance, $\Delta (\Sigma E_T)$, has  smaller systematic uncertainty than energy sums for the correlated peaks themselves, as most sources of systematic errors cancel out as discussed above.
A non-zero value of $\Delta (\Sigma E_T)$ could result from  QGP medium effects and/or the known $k_{T}$-effects  characteristic for back-to-back partons~\cite{phenix_ppg029}. To disentangle these two contributions the measured  energy imbalance for di-jets from central Au+Au events is compared with  the corresponding value obtained from  d+Au data with the same trigger/associated particle $p_{T}$ cuts. 
The energy imbalance for symmetric trigger pairs~\cite{short_2plus1_paper} was found to be $1.59\pm 0.19$~GeV/$c$ in central Au+Au collisions, similar to the value of $1.65 \pm 0.39$~GeV/$c$ reported for the minimum bias d+Au data. 
This value is also close to the initial state $k_{T}$ effects of $\sim$1.6 GeV/$c$ estimated in di-jet correlation \cite{renk_kt} and disfavors additional partonic  energy loss into the medium for these Au+Au data.
The detailed results for our new measurement for the asymmetric trigger pairs are summarized in Tables \ref{table_sum_etpt_dau} and  \ref{table_sum_etpt_auau}.
In each table,  only the statistical uncertainties are listed for each variable. The statistical uncertainties on mean values of trigger  $E_T$ or $p_T$ are negligible.
\begin{table*}[thb]
\caption{Jet energy estimate for the near- and away-sides of back-to-back triggered correlations from 200 GeV  d+Au data. All units are in GeV or GeV/$c$. For both primary trigger selections listed, the secondary trigger is required to have transverse momentum  $p_T \in [4,10]$ GeV/$c$.}
\begin{tabular*}{0.99\textwidth}{c|c|c|c|c|c|c}
\hline
& & & & & &\\ 
 Trigger 1   & $\langle E_T \rangle^{\rm{T1}}$(GeV)  & $\langle p_T \rangle^{\rm{T2}}$(GeV/$c$)  & Assoc $p_T$(GeV/$c$)   & $\sum p_T^{\rm{assoc},\rm{ near}}$(GeV/$c$)    &  $\sum p_T^{\rm{assoc}, \rm{away}}$(GeV/$c$)     & $\Delta (\Sigma E_T)$(GeV)  \\
& & & & & & \\
\hline
			&      &     & [1.0,10.0] &  4.09$\pm$0.06  & 4.68$\pm$0.06 &   2.74$\pm$0.09 \\
$E_T\in$[8,10]GeV  & 8.89 & 5.56& [1.5,10.0] &  3.49$\pm$0.06  & 3.94$\pm$0.06 &   2.87$\pm$0.08 \\
            &      &     & [2.0,10.0] &  2.95$\pm$0.06  & 3.26$\pm$0.06 &   3.02$\pm$0.08 \\
            &      &     & [2.5,10.0] &  2.47$\pm$0.06  & 2.62$\pm$0.05 &   3.18$\pm$0.08 \\
\hline
			&      &	  &[1.0,10.0] &  4.14$\pm$0.07  & 5.75$\pm$0.08 &  4.26$\pm$0.11 \\
$E_T\in$[10,15]GeV & 11.74& 5.86 &[1.5,10.0] &  3.59$\pm$0.07  & 4.88$\pm$0.08 &  4.58$\pm$0.10 \\
			&      &      &[2.0,10.0] &  3.09$\pm$0.07  & 4.02$\pm$0.07 &  4.95$\pm$0.10 \\
		 	&      &      &[2.5,10.0] &  2.65$\pm$0.07  & 3.32$\pm$0.07 &  5.20$\pm$0.10 \\
\hline
\end{tabular*}
\label{table_sum_etpt_dau}
\end{table*}

\begin{table*}
\caption{Jet energy estimate for the near- and away-sides of back-to-back triggered correlations from 200 GeV central  Au+Au collisions. All units are in GeV or GeV/$c$. For both primary trigger selections listed, the secondary trigger is required to have transverse momentum  $p_T \in [4,10]$ GeV/$c$.}
\begin{tabular*}{0.99\textwidth}{c|c|c|c|c|c|c}
\hline
& & & & & &\\ 
 Trigger 1   & $\langle E_T \rangle^{\rm{T1}}$(GeV)  & $\langle p_T \rangle^{\rm{T2}}$(GeV/$c$)  & Assoc $p_T$(GeV/$c$)   & $\sum p_T^{\rm{assoc}, \rm{near}}$(GeV/$c$)    &  $\sum p_T^{\rm{assoc}, \rm{away}}$(GeV/$c$)     & $\Delta (\Sigma E_T)$(GeV)  \\
& & & & & & \\
\hline
		   &      &       & [1.0,10.0] & 4.68$\pm$0.27    & 4.82$\pm$0.25  & 3.45$\pm$0.37 \\
$E_T\in$[8,10]GeV & 8.87 &  5.27 & [1.5,10.0] & 3.81$\pm$0.20    & 3.86$\pm$0.17  & 3.55$\pm$0.26 \\
           &      &       & [2.0,10.0] & 3.16$\pm$0.14    & 2.94$\pm$0.12  & 3.82$\pm$0.19 \\
           &      &       & [2.5,10.0] & 2.61$\pm$0.11    & 2.13$\pm$0.09  & 4.08$\pm$0.15 \\
\hline
			& 		& 		& [1.0,10.0] & 4.12$\pm$0.29  & 4.96$\pm$0.24  & 5.21$\pm$0.38 \\
$E_T\in$[10,15]GeV & 11.76 & 5.70  & [1.5,10.0] & 3.55$\pm$0.21  & 4.00$\pm$0.17  & 5.61$\pm$0.26 \\
			& 		& 		& [2.0,10.0] & 3.19$\pm$0.14  & 3.10$\pm$0.12  & 6.15$\pm$0.19 \\
			& 		& 		& [2.5,10.0] & 2.78$\pm$0.11  & 2.27$\pm$0.10  & 6.56$\pm$0.15 \\
\hline
\end{tabular*}
\label{table_sum_etpt_auau}
\end{table*}

\begin{figure}[!thb]
\includegraphics[width=0.45\textwidth]{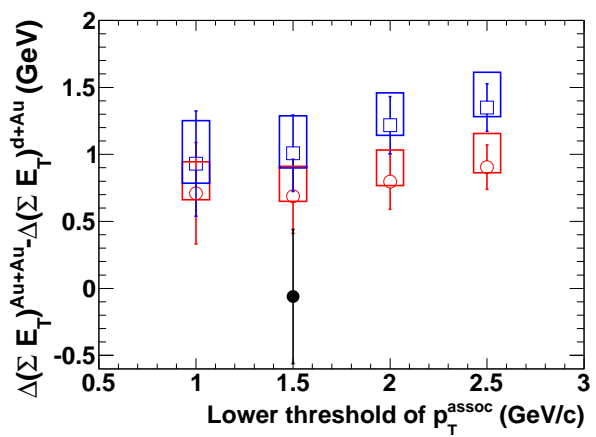}
\caption{(Color online) The relative di-jet energy imbalance estimate $\Delta (\Sigma E_T)^{\rm{Au+Au}} - \Delta (\Sigma E_T)^{\rm{d+Au}} $.  The open circles show results for primary triggers in [8,10]~GeV, the open squares - [10,15]~GeV. The solid data point is the result for symmetric trigger pairs from previous STAR paper~\cite{short_2plus1_paper}. The x-axis shows lower threshold of the transverse momentum selection for associated hadrons.\\}
\label{figure_delta_sigma_energy_auau_dau}
\end{figure}

In Fig.~\ref{figure_delta_sigma_energy_auau_dau} the relative energy imbalance between central Au+Au and  d+Au data (e.g. $\Delta (\Sigma E_T)^{\rm{Au+Au}}-\Delta (\Sigma E_T)^{\rm{d+Au}}$) is shown for both the symmetric (solid symbol) and asymmetric trigger cases (open symbols).
The errors shown include both statistical and systematic uncertainties discussed in the previous section.

The observed relative energy imbalance $\Delta (\Sigma E_T)^{\rm{Au+Au}} - \Delta (\Sigma E_T)^{\rm{d+Au}} $ between Au+Au and d+Au for the asymmetric trigger pairs studied is consistent with non-zero values.
The relative imbalance is consistently higher for all associated hadron selections for more asymmetric triggers,  as expected if larger asymmetries are related to longer paths traversed and higher energy loss.
The measured values of relative energy imbalance are significantly smaller than the theoretical prediction for in-medium energy deposition based on path-dependent energy loss~\cite{renk_kt}. In this model approximately 3~GeV is expected for our asymmetric case. 
The ``2+1'' results for both symmetric and  asymmetric trigger pairs seem to favor much stronger surface bias than expected in theory and thus point to stronger path-length dependence of energy loss. 
Also, as previously discussed, any direct-$\gamma$  contamination to our primary trigger sample would reduce the observed energy imbalance, leading to even greater discrepancy between the data and the theory.
We note, however, that if part of energy deposited to the medium by the traversing parton produces hadrons that remain within the angular selection of our jet-peaks,  it will not be observed in the relative imbalance variable.
Thus it is possible, that we only measure part of the lost energy that got redistributed to hadrons that are softer than those considered in this study, or that were radiated at larger angles than those included in our jet-cone selection.  The dependence of observed energy imbalance on the associated hadron $p_T$ seems to confirm this idea.
The relative imbalance is found largest for the highest associated hadron $p_T$ bin studied, reaching the value 
$\sim$1.5 GeV/$c$, and decreasing for the softer associated hadrons to approximately $\sim$1~GeV/$c$.
This trend can indicate  that the jet fragmentation  is shifted to the lower $p_T$ regime (softening), or that the lower energy  fragmentation products are at larger angles from the jet axis (broadening). The later is less likely due to consistent jet-like peak widths observed in Au+Au and d+Au events.
In either of the two cases, in the limit of associated hadron  $p_T$ approaching zero, the value of relative $\Delta (\Sigma E_T)^{\rm{Au+Au}} - \Delta (\Sigma E_T)^{\rm{d+Au}} $ consistent with zero would indicate that full  jet energy is recovered. Extrapolating our data to very low momenta,  albeit large uncertainties, seems consistent with this scenario. 

\begin{figure}[!thb]
\includegraphics[width=0.45\textwidth]{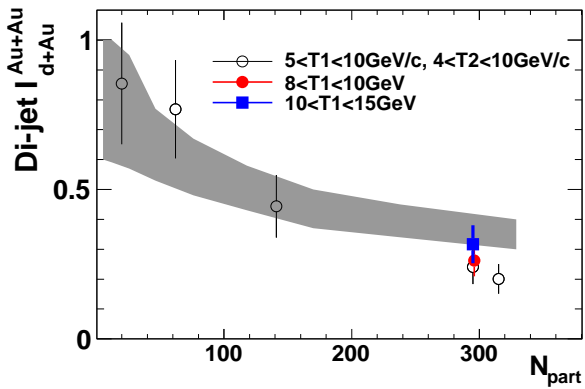}
\caption{(Color online) Conditional di-jet survival probability measured through  di-jet relative suppression rates $I_{\rm{d+Au}}^{\rm{Au+Au}}$ as a function of centrality. The circles show data from \cite{short_2plus1_paper}. Filled symbols show new measured ratios for the asymmetric trigger pairs with primary triggers of [8,10]~GeV and [10,15]~GeV. The band shows the expectation for di-jet surface emission (``corona'' only)  rates from \cite{short_2plus1_paper}.}
\label{figure_di_trigger_rate_auau_dau}
\end{figure}

Additional information could be obtained from the relative di-jet production rates. For the  surface-emission scenario the jets and di-jets production rates are determined by the surface volume of the fireball. 
In simplest implementation of this scenario the medium in heavy-ion collisions is modeled consisting of a  completely opaque core (leading to full jet absorption) surrounded by a permeable corona (with no jet-medium interactions).
In such a model a  parton (jet) will survive with no energy loss or modifications if the underlying  hard-scattering happened in the corona region and the scattered parton did not  pass through the core.
A Monte Carlo Glauber model based on such a model was compared to the early di-jet production rates in~\cite{short_2plus1_paper}.
The size of the permeable corona  was first determined from tuning to the measured single charged hadron $R_{\rm{AA}}$ results from RHIC.
It was then  used to predict the di-jet  survival probability for the back-to-back triggers. 
To minimize systematic uncertainties related to the determination of the number  of binary collisions ($N_{\rm{coll}}$) and the number of participating nucleons ($N_{\rm{part}}$) in the Glauber model, the double ratios measuring conditional di-jet survival probability  were used. 
Such a double ratio, $I^{ \rm{Au+Au} }_{ \rm{ d+Au } } = \frac{ R_{\rm{AA}}^{\rm{\ di-jet \ triggers}} } { R_{\rm{AA}}^{\rm{\ single \ triggers}} }$, constructed from nuclear modification factors ($R_{\rm{AA}}$, the binary scaled ratios of the observables in Au+Au and d+Au collisions), reflects any changes in probability to find an away-side trigger for each primary trigger observed in Au+Au data relative to d+Au. 
%
%
%
In the earlier work the value $I_{\rm{d+Au}}^{\rm{Au+Au}}=0.20 \pm 0.05$ for the symmetric trigger pairs was found qualitatively consistent with the estimates of a simplistic core/corona model~\cite{short_2plus1_paper}.
The di-jet survival probability was calculated the same way in this work for the  new asymmetric trigger data.
These  additional new points for the 0-20\% central Au+Au data are included in Fig.~\ref{figure_di_trigger_rate_auau_dau} which reproduces the plot from the previous work.
We find the conditional survival rate $I_{\rm{d+Au}}^{\rm{Au+Au}}$ for the asymmetric trigger pairs to be  $0.26 \pm 0.05$ for $E_{T}^{\rm{T1}}\in[8,10]$~GeV, and $0.32 \pm 0.06$ for $E_{T}^{\rm{T1}}\in[10,15]$~GeV.
We note that the dominant source of systematic uncertainties comes from  $N_{bin}$ calculation in the Glauber model, which is then fully correlated between the data points of the same centrality bin (new values) and largely correlated between the values for central collisions between the data from different runs. So the slight increase in di-jet conditional survival with increasing trigger pair asymmetry points to higher in-medium path-lengths, which is consistent with our conclusions based on the relative energy imbalance.
In other words, the new values of $I_{\rm{d+Au}}^{\rm{Au+Au}}$ indicate that jets triggered by those more asymmetric back-to-back pairs  are  emerging from deeper within the ``core'' of the medium compared to a more symmetric pair  selection.

%
%
%
%
%
%


\section{Summary}

Jet-medium interactions were studied via a ``2+1'' multi-hadron correlation technique. 
For both symmetric and asymmetric trigger pairs, the distributions of associated hadrons and their spectra show no strong shape modifications from d+Au to central Au+Au collisions on both near- and away- trigger sides.
This is in contrast to the  di-hadron correlation results with respect to a single high-$p_T$ trigger. In addition, no evidence of the near-side ``ridge''~\cite{star_ridge} or the away-side ``shoulders''~\cite{phenix_corr} was observed in these ``2+1'' correlations. 
The relative total transverse momentum imbalance was measured as an excess in the difference between the sum of the momentum (or energy) for hadrons attributed to the near-side and away-side jet-like peaks in Au+Au with respect to the reference d+Au data.
The relative imbalance of Au+Au over d+Au has been shown to be non-zero for the asymmetric trigger pair selections, and increasing with the asymmetry of the trigger pair. 
This is consistent with expected medium effects for partons probing deeper within the medium. This challenges the simplistic implementation of the full absorption ``core/corona'' model, which captured well the centrality trend of di-jet production rates for the symmetric triggers.
The relative imbalance is found largest for the highest associated hadron $p_T$ bin studied, reaching the value 
$\sim$1.5 GeV/$c$, and decreasing for the softer associated hadrons. This trend indicates that the energy missing from the away-side peak at higher associated momenta is converted into softer hadrons. 
The measured relative imbalance for all bins is  
less than the theoretical predictions for total in-medium energy deposition for such trigger pairs of  $\sim$3 GeV/$c$ in the path-length dependent energy loss model~\cite{renk_kt}. We note, however, that if part of energy deposited to the medium by the traversing parton produces hadrons that remain within the angular selection of our jet-peaks,  it will not be observed in the relative imbalance variable.
%
%

We thank the RHIC Operations Group and RCF at BNL, the NERSC Center at LBNL and the Open Science Grid consortium for providing resources and support. This work was supported in part by the Offices of NP and HEP within the U.S. DOE Office of Science, the U.S. NSF, the Sloan Foundation, CNRS/IN2P3, FAPESP CNPq of Brazil, Ministry of Ed. and Sci. of the Russian Federation, NNSFC, CAS, MoST, and MoE of China, GA and MSMT of the Czech Republic, FOM and NWO of the Netherlands, DAE, DST, and CSIR of India, Polish Ministry of Sci. and Higher Ed., National Research Foundation (NRF-2012004024), Ministry of Sci., Ed. and Sports of the Rep. of Croatia, and RosAtom of Russia.





\end{document}